\documentclass[draft]{agujournal2019}
\usepackage{url} 
\usepackage{lineno}
\usepackage[finalnew]{trackchanges} 
\usepackage{soul}
\draftfalse

\journalname{Journal of Geophysical Research}

\begin{document}

\addeditor{Adam}
\addeditor{Imai}
\addeditor{Louis}

%
%


\title{Generation mechanism and beaming of Jovian nKOM from 3D numerical modeling of Juno/Waves observations}

%
%




\authors{A. Boudouma\affil{1}, P. Zarka\affil{1,2}, C.K. Louis\affil{1,3}, C. Briand\affil{1}, M. Imai\affil{4}}


\affiliation{1}{LESIA, Observatoire de Paris, CNRS, PSL, Sorbonne Université, Université Paris Cité, Meudon, France}
\affiliation{2}{Observatoire Radioastronomique de Nançay, ORN, Observatoire de Paris, CNRS, PSL, Université d'Orléans, Nançay, France}
\affiliation{3}{School of Cosmic Physics, DIAS Dunsink Observatory, Dublin Institute for Advanced Studies, Dublin, Ireland}
\affiliation{4}{Department of Electrical Engineering and Information Science, National Institute of Technology (KOSEN), Niihama College, Niihama, Japan}




\correspondingauthor{Adam Boudouma}{adam.boudouma@obspm.fr}




\begin{keypoints}
\item We developed a 3D modeling method to test the generation mechanism and beaming of plasma emissions from Jupiter's inner magnetosphere.

\item nKOM occurrence distribution is reproduced with plasma emissions at $f_{pe}$ beamed in the opposite direction of their frequency gradient.

\item nKOM is compatible with O-mode at high latitudes and X-mode at low latitudes, from radio sources distributed near the centrifugal equator.
\end{keypoints}

%
%

%
%


\begin{abstract}
The narrowband kilometric radiation (nKOM) is a Jovian low-frequency radio component identified as a plasma emission produced in the region of the Io plasma torus. Measurements from the Waves instrument onboard the Juno spacecraft permitted to establish the distribution of nKOM occurrence and intensity as a function of frequency and latitude.
We have developed a 3D geometrical model that can simulate at large scale the plasma emissions occurrence observed by a spacecraft based on an internal Jovian magnetic field model and a diffusive equilibrium model of the plasma density in Jupiter's inner magnetosphere. With this model, we propose a new method to discriminate the generation mechanism, wave mode, beaming and radio source location of plasma emissions. Here, this method is applied to the study of the nKOM observed from all latitudes by the Juno/Waves experiment to identify which conditions reasonably reproduce the observed occurrence distribution versus frequency and latitude. The results allow us to exclude the two main nKOM models published so far, and to show that the emission must be produced at the local plasma frequency and beamed along its local gradient in the direction of decreasing frequencies. We also propose that depending on its latitude, Juno observes two distinct kinds of nKOM: the low frequency nKOM in ordinary mode at high latitudes and high frequency nKOM on extraordinary mode at low latitudes. Both radio source locations are found to be distributed near the centrifugal equator from the outer edge to the inner edge of the Io plasma torus.
\end{abstract}

\section*{Plain Language Summary}

This paper investigates a specific type of Jupiter's natural radio emissions called ``narrowband kilometric radiation" (nKOM). The nKOM is produced within the region of Jupiter's Io plasma torus. Using data collected by the Waves instrument on the Juno spacecraft, we analyze when and how the nKOM occurs across different frequencies and latitudes. To better understand these radio emissions, we have developed a 3D model that simulates the occurrence of nKOM. 
Additionally, we introduce a novel methodology for determining key characteristics of nKOM, including its generation mechanism, propagation mode, beaming direction, and the source location of the radio emissions. Applying his model to Juno's observations, we found that the previously proposed theoretical models for nKOM generation do not match the observed data. Instead, the study suggests that nKOM is generated at the local plasma frequency and is beamed along its local gradient, in the direction of decreasing frequencies. The study also revealed, two distinct types of nKOM depending on Juno's latitude: low-frequency nKOM in ordinary mode at high latitudes and high-frequency nKOM in extraordinary mode at low latitudes. Both types of nKOM are found to originate near the Jupiter's centrifugal equator.

\section{Introduction}

A large variety of radio emissions are produced below 40 MHz in the inner Jovian magnetosphere. Only the decameter emissions are accessible to ground-based observations due to the Earth's ionospheric cutoff at 10 MHz. 
But observations from the two Voyager spacecraft extended Jupiter's radio spectrum to the hectometer and kilometer wavelength ranges \cite{warwick1979a,warwick1979b}. Two components were discovered in the kilometer range: a broadband ($\sim$10 kHz to $\sim$1 MHz) sporadic emission called bKOM \cite{desch1980}, and a narrowband smoother one restricted to frequencies around 100 kHz called nKOM \cite{kaiser1980}. 
Both kilometer wave components display a minimum of occurrence for an observer near Jupiter's equator, more marked for bKOM \cite{kurth1980,daigne1986,louis2021}. 
This led \citeA{jones1980,jones1986,jones1987} to propose for both bKOM and nKOM a generation theory based on the linear conversion of Z-mode electrostatic waves at the upper hybrid resonance frequency $f_{uh}=(f_{pe}^2+f_{ce}^2)^{1/2}$ (with $f_{pe}$ the electron plasma frequency and $f_{ce}$ the electron cyclotron frequency) into O-mode waves.
But studies of Voyager data soon showed that for bKOM, polarization is consistent with X-mode waves and occurrence is modulated at Jupiter's rotation period (system III, $\sim9.925$h) \cite{desch1980,leblanc1985}, while nKOM is more consistent with an O-mode emission and is modulated at a variable period $\sim$3\% to 5\% longer than Jupiter's system III rotation period \cite{kaiser1980,daigne1986}.
Next, radio goniopolarimetric (i.e. direction-finding + polarization) measurements by the Ulysses spacecraft showed unambiguously that bKOM originates from the high-latitude auroral region \cite{ladreiter1994} whereas nKOM originates from the vicinity of Io's plasma torus (IPT, extending from slightly under 6 $R_J$ to $\sim$ 12 $R_J$, with $R_J~=~71492$ km the Jovian radius) \cite{reiner1993}. This scenario is supported by Juno measurements \cite{imai2017}.
These results identified the bKOM as an X-mode auroral component produced by the electron cyclotron maser mechanism near the local $f_{ce}$ \cite{leblanc1985,wu1985,zarka1998,treumann2006}, and they confirmed the early hypothesis that nKOM was a low-latitude plasma emission (near $f_{pe}$ or $2~f_{pe}$) mainly produced on the O-mode near the outer edge of the IPT \cite{kaiser1980,carr1983,daigne1986, reiner1993}.

Beyond these qualitative conclusions, there is no consensus about the detailed conditions for generation of nKOM, namely its frequency (near the local $f_{pe}$ \cite{jones1980,jones1986,jones1987} or $f_{uh}$, or its first harmonic \cite{fung1987}), its beaming (along the local magnetic field $\bf{B}$, the electron density gradient $\nabla n_e$, or at some specific angle relative to these vectors), and its dependence on the density gradient strength $||\nabla n_e||$, the magnetic field amplitude $|\bf{B}|$, or the angle between $\nabla n_e$ and $\bf{B}$. Despite O-mode being seemingly dominant, as nKOM have also been observed in X-mode \cite{reiner1993}, its favored mode of emissions and their sources locations, remains to be confirmed.

In the absence of wave and particle measurements inside nKOM radio sources, that would allow us to answer the above questions as was done for auroral radio sources \cite<see e.g.>[]{zarka1998, louarn2017, louarn2018, louis2020, collet2023}, we investigate here if large-scale modeling can reproduce the observed statistical distribution of nKOM occurrence versus frequency and latitude \cite{louis2021}, in order to provide strong constraints on its conditions of emission and thus answer at least part of the above questions.

In section 2, we detail key aspects of the theories proposed for nKOM generation, and the hypotheses and parameters that will be tested via our modeling. In section 3, we present the observational data that our modeling intends to reproduce, for testing the relevance of each theory and hypothesis. Section 4 describes the plasma and field models used and our modeling framework. The results are presented in section 5, and discussed in section 6 that also gives further perspectives to this work.

\section{Theories}\label{sec:theory}

As mentioned in the introduction, observations from Voyager, Ulysses and Juno suggest that nKOM is produced in or near the IPT. 
In that region $f_{pe} \ge f_{ce}$ or $f_{pe} \sim f_{ce}$, favouring mode conversion mechanisms \cite{melrose2017}.
The literature proposes two classes of theories for the generation of nKOM, from which we define four scenarios below.

The first theory, developed by \citeA{jones1980, jones1986, jones1987}, is based on the radio window theory \cite{budden1966, budden1971, jones1976, budden1980} applied to the Jovian magnetosphere to explain the latitudinal beaming of the nKOM observed by Voyager. They predict Z-mode electrostatic waves, to produce Cerenkov radiation when $f_{pe} < f < f_{uh}$ that cannot escape the Io plasma torus. If the Z-mode wave propagates toward the decreasing plasma density, it is either absorbed or reflected at the upper-hybrid resonance $f \sim f_{uh}$. If the Z-mode wave propagate toward the increasing plasma density, it is absorbed when reaching the plasma frequency $f \sim f_{pe}$. However, if the Z-mode and the O-mode refractive indexes are close when $f \sim f_{pe}$, the Z-mode is able to convert into O-mode through the so-called ``radio window". The process requires density gradients $\nabla n_e$ to be weak enough for the radio window to occur according to the ray theory approximations. They also suggest Z-mode quasi-perpendicular propagation with respect to the magnetic field $\bf{B}$ (which is equivalent to quasi-perpendicularity between $\bf{B}$ and $\nabla n_e$) to full-fill upper-hybrid resonance and maximize the mode conversion efficiency. The O-mode radio wave, first generated along $\nabla n_e$, is finally beamed in two opposite directions making angles $\beta~=~\arctan (f_{pe}/f_{ce})^{1/2}$ and $\pi~-~\beta$ with respect to $\bf{B}$ in the plane defined by ($\nabla n_e$, $\bf{B}$). This will be our scenario~\#1.

The theory of \citeA{fung1987} conversely suggests that nKOM is produced at $2f_{uh}$ by nonlinear coupling of two electrostatic waves each at $f_{uh}$, propagating through density inhomogeneities at the outer periphery of the IPT. Emission is favoured at places where the density gradient $\nabla n_e$ is weak and makes a specific angle ($\sim47^\circ$) with the local $\bf{B}$. The theory predicts both O-mode and X-mode emissions, beamed in a plane nearly perpendicular to $\bf{B}$. This will be our scenario~\#2.

The radio window theory has been updated by \citeA{budden1986}, removing the ray theory approximation, as observations have shown that larger density gradients are likely to be present in the potential plasma emissions generation regions. Applied to the Jovian magnetosphere it can predict the conversion of Z-mode into O-mode and X-mode if the density gradient is strong enough. Following the results from \citeA{budden1986}, the beaming angle $\beta$ from \citeA{jones1980} might not be relevant to describe the beaming of O-mode nKOM when the density gradient is strong. Thus, we will also consider a scenario~\#3 where emission is produced (through mode conversion) at the local $f_{pe}$ along its local gradient $- \nabla f_{pe}$ and, for completeness, a similar scenario~\#4 with emissions produced at the local $f_{uh}$ along its local gradient ($- \nabla f_{uh}$). These four nKOM generation scenarios are summarized in Table \ref{tab:cases}. A more detailed version of this table is provided in the appendix (Table \ref{tab_appendix:cases}).

\section{Data}\label{sec:data}

\begin{figure}[!ht]
\centering
\noindent\includegraphics[width=\textwidth]{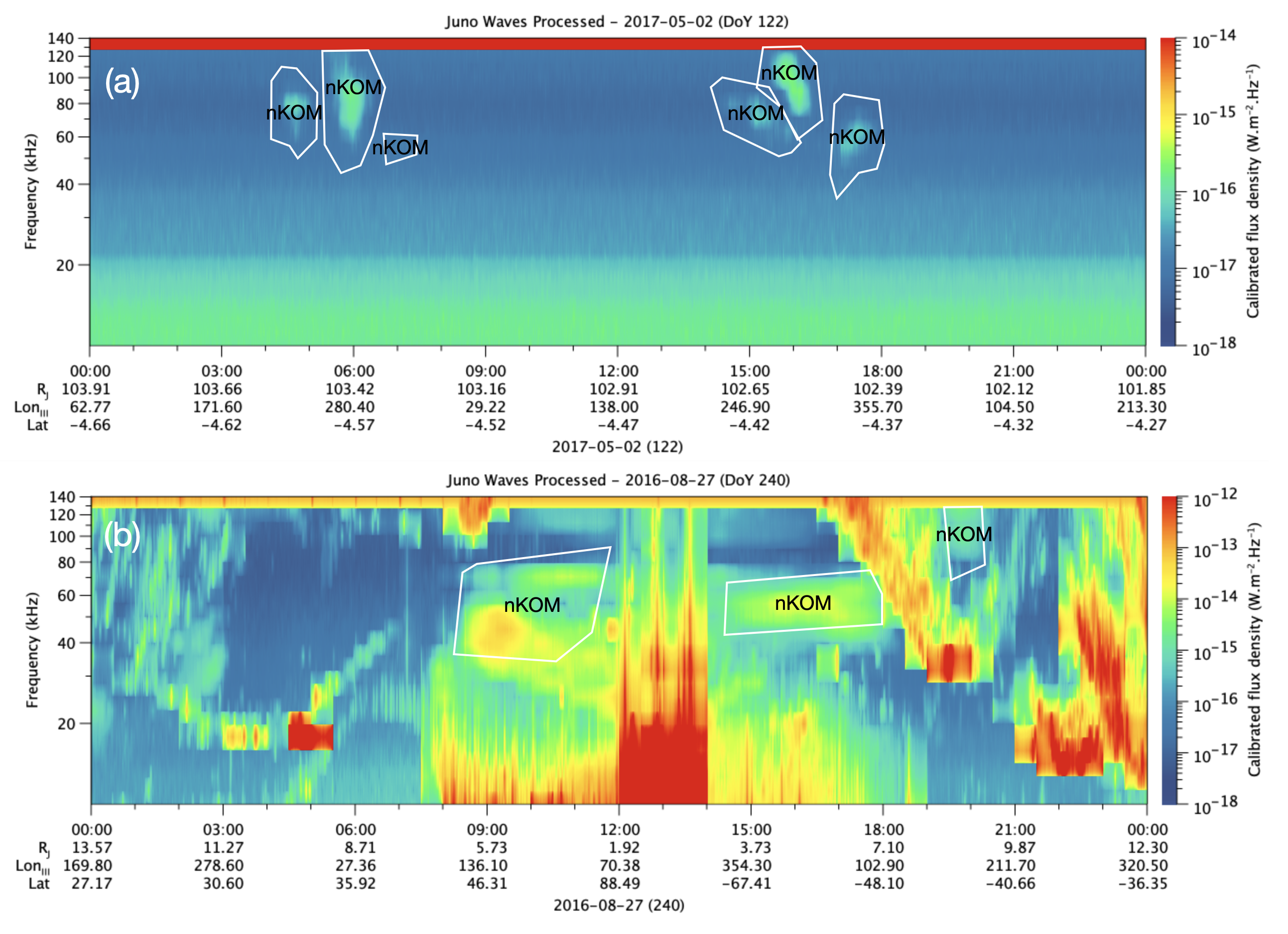}
\caption{Two 24-hour dynamic spectra of calibrated Juno/Waves data \cite{louis2021} observed (a) from low latitudes (day-of-year 2017-122, i.e. 05/02/2017) and (b) around the first perijove PJ1 (day-of-year 2016-240, i.e. 27/08/2017). The white boxes correspond to the cataloged nKOM components with their labels displayed in black \cite{louis2021_catalog}. The horizontal axis corresponds to the Juno observations time period, radial distance (R$_{\mathrm{j}}$), system III longitude (Lon$_{\mathrm{III}}$) and jovicentric latitude (Lat). The vertical axis represent the Juno/Waves frequency channels from 10 kHz to 141 kHz, distributed on a logarithmic scale with a ratio $\times 1.12$ between consecutive channels. The saturated band at 140 kHz correspond to the channel 61 of the HFR-low receiver that has a poor sensitivity. The two color bars represent the calibrated flux densities measured by the Waves experiment at different times. The intensity scales are adjusted to highlight the nKOM occurrences.}
\label{fig:spdyn_juno}
\end{figure}

In Voyager \cite{kaiser1980,daigne1986}, Galileo \cite{louarn1998}, Cassini \cite{zarka2021} and Juno \cite{imai2017,louis2021} data, nKOM is detected as smooth patches lasting for one to a few hours, that can reoccur every 10.3h to 10.5h during several days (cf. Figure \ref{fig:spdyn_juno}). This periodicity is consistent with the plasma motion in the outer parts of the IPT, whose inertia causes slight subcorotation relative to Jupiter's magnetic field rotation at the system III period. 
\citeA{louarn1998,louarn2000} further proposed a scenario explaining intermittent activity periods of 3-4 days by centrifugal ejections of plasma from the IPT, exciting locally nKOM and also causing brightenings of auroral emissions (bKOM, hectometer emission). 
An effect of the solar wind causing variable magnetospheric compressions is not excluded \cite{nozawa2006, louis2023}, although it is expected to be stronger in the outer magnetosphere \cite{zarka2021} than in the IPT.
This intermittency of nKOM makes difficult to characterize it on an event by event basis. We thus focused on statistical constraints.

\begin{figure}[!ht]
\centering
\noindent\includegraphics[width=0.65\textwidth]{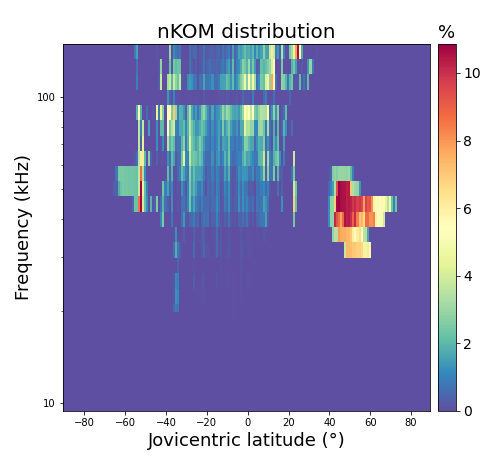}
\caption{nKOM occurrence probability versus frequency and latitude derived from the first 3 years of Juno/Waves low-frequency observations around Jupiter by \citeA{louis2021}. 
Latitude bins are 1° wide and frequencies correspond to Juno/Waves frequency channels from 10 kHz to 141 kHz, distributed on a logarithmic scale with a ratio $\times 1.12$ between consecutive channels. 
This figure is based on the same data  as Fig. S4a from \citeA{louis2021}, except that no smoothing has been applied.}
\label{fig:ckl}
\end{figure}

Based on 6 months of Cassini radio observations from low Jovian latitudes, \citeA{zarka2004} found that nKOM covers the range $\sim60-160$ kHz (very different from the more sporadic bKOM that covers the range $\sim23-400$ kHz in that same data set). Statistics over 3 years of Juno radio observations from all latitudes (from 9 April 2016 to 24 June 2019) allowed \citeA{louis2021} to extend this range down to 20-30 kHz and to characterize the variations of nKOM occurence (and intensity, but we only use the occurrence here) with the latitude of the observer (cf. Figure \ref{fig:ckl}). Juno could not precise the upper frequency limit of nKOM because its radio instrument has a very limited sensitivity in the range 140 kHz--3 MHz. The time-averaged distribution of Figure \ref{fig:ckl} is asymmetrical between the two Jovian hemispheres. Above 60 kHz, nKOM occurrence peaks at $-40^\circ \pm 15^\circ$ in the South and $+15^\circ \pm 15^\circ$ in the North, with a minimum of occurrence about $-15^\circ$, and a gap around $+30^\circ$ below $\sim$100 kHz. At lower frequencies, one observes extensions of nKOM occurrence down to $-70^\circ$ latitude around 50 kHz, and a very strong peak up to $+75^\circ$ latitude at $30-50$ kHz. 

We display on Figure \ref{fig:spdyn_juno} representative examples of the Juno/Waves nKOM observations at two geometrical configurations: (a) when Juno is far from Jupiter and at low latitudes and (b) when Juno is close to Jupiter and at high latitudes (near its perijove). Panel (a) of Figure \ref{fig:spdyn_juno} displays two occurrences of nKOM events separated by $\sim$ 10h, while panel (b) displays two distinct nKOM events observed each for a longer period of time but without periodicity. The corresponding radio sources may be wider in longitudes than panel (a), or the occurrence is here dominated by the variation of position of Juno along its perijove orbit. Furthermore, when Juno is (a) at low latitudes, the nKOM is observed for frequencies roughly $>40$ kHz, while when Juno is (b) at high latitudes, the nKOM is observed at significantly lower frequencies, with a different frequency range for the nKOM observed before the perijove (30-50 kHz) and the nKOM observed after the perijove (40-60 kHz).

Figures \ref{fig:spdyn_juno} and  \ref{fig:ckl} suggest that nKOM is produced at specific locations in the IPT. The variable frequency ranges observed at low and high latitudes by Juno, should provide information on the large-scale beaming and distribution of the radiosource inside the IPT.

\section{Modeling}\label{sec:modeling}

Our purpose is to derive significant constraints on nKOM generation from the modeling of its large-scale time-averaged latitude-frequency distribution of Figure \ref{fig:ckl}.
We will thus apply the conditions defined in each scenario to a grid of modeled plasma density and magnetic field values in Jupiter's inner magnetosphere, synthesize the frequency-latitude occurrence distribution of nKOM detected along the trajectory of Juno between 9 April 2016 and 24 June 2019, and compare it to the latitude versus frequency distribution of the observations displayed on Figure \ref{fig:ckl}.

\subsection{Plasma Density \& Magnetic Field Models}\label{sec:model}

To model the plasma density in Jupiter's inner magnetosphere and IPT, we use the diffusive density model of \citeA{imai2016} that computes the densities of the main torus species (electrons, many positive ions of $S$ \& $O$, $H^+$ and $Na^+$) for radial distances between 4 and 13 $R_J$ and jovicentric latitudes below $80^\circ$. This model is based on time-averaged plasma (density and temperature) observations of Voyager 1 and 2 as a function of the radial distance to Jupiter \cite{bagenal1994,moncuquet2002}. The plasma density is then predicted in a meridian plane by extrapolating the radial profiles along Jupiter's magnetic field lines using the VIP4 magnetic field model plus current sheet \cite{connerney1998}. For 3D representation of the plasma density, cylindrical symmetry around the jovicentrifugal axis is assumed.
The \citeA{imai2016} model has been preferred to the model by \citeA{divine1983} because it computes the density in a more consistent way out of the region where measurements were made.

The VIP4 magnetic field model (based on Voyager and Pioneer magnetic field measurements plus infrared observations of the Io flux tube footprints) is limited to order 4. More accurate models have been built from Juno magnetic measurements. The most recent one, JRM33 \cite{connerney2022}, includes spherical harmonics up to order 13. But to describe the magnetospheric magnetic field between 4 and 13 $R_J$, the VIP4 plus current sheet model is still adequate, because at these distances the influence of higher order terms in the magnetic field becomes negligible, and it is fully consistent with the \citeA{imai2016} plasma model.

\begin{figure}[!ht]
\centering
\noindent\includegraphics[width=0.75\textwidth]{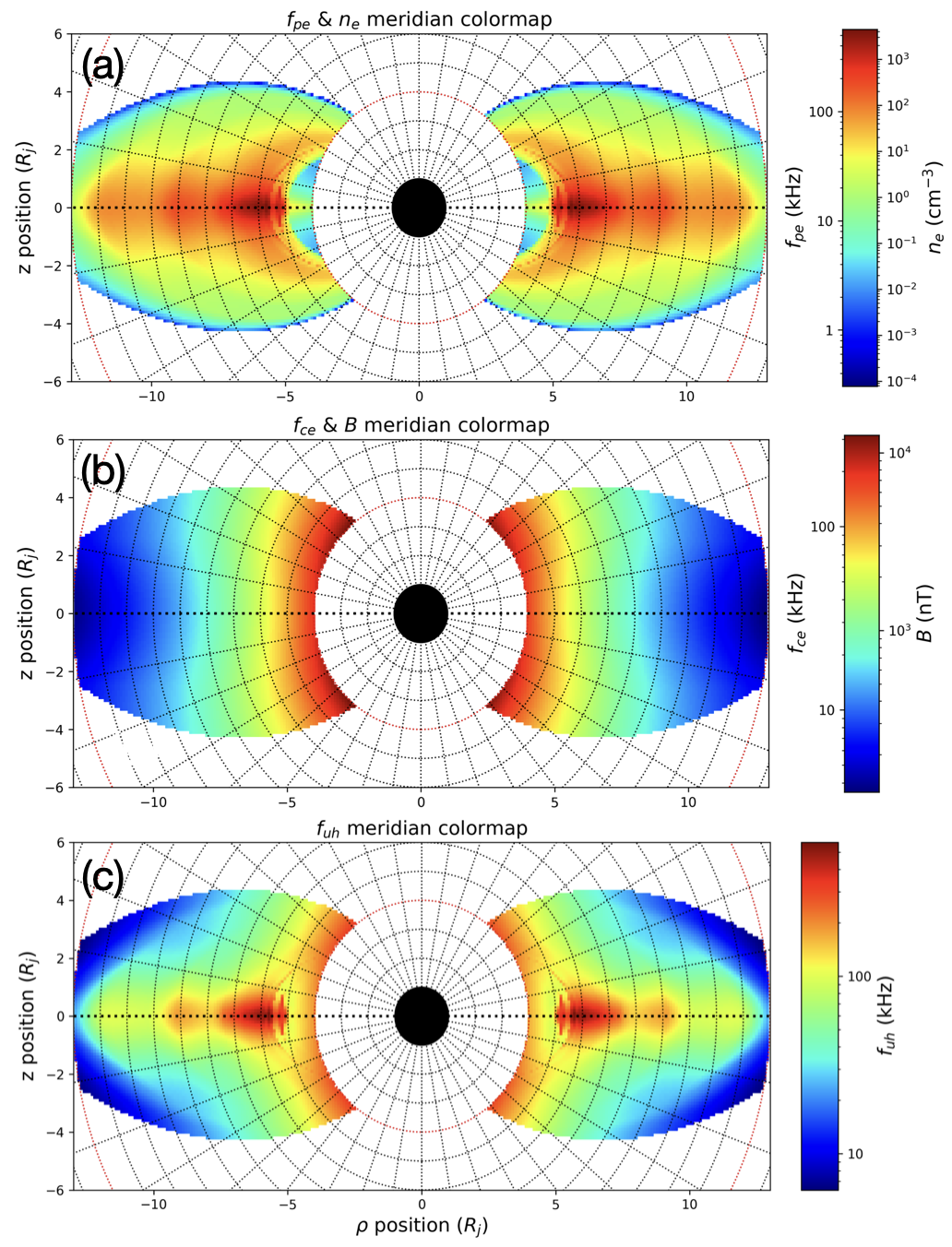}
\caption{Meridian maps of (a) the plasma frequency $f_{pe}~=~\sqrt{n_e e^2/ 4 \pi^2 m_e \epsilon_0}$ computed from the electron density $n_e$ of the diffusive density model of \citeA{imai2016}, (b) the electronic cyclotron frequency $f_{ce}~=~eB/2\pi m_e$ computed from the magnetic field amplitude $B$ from the VIP4 magnetic field model \cite{connerney1998} and (c) the upper hybrid frequency $f_{uh}~=~\sqrt{f_{pe}^2 + f_{ce}^2}$. The meridian plane here is where the jovicentric, jovicentrifugal and jovimagnetic equators are aligned (defined by the system III longitude $\varphi~=~200.769^{\circ}$). The vertical axis is the cartesian coordinate $z$. The horizontal axis is the radial cylindrical coordinate $\rho$ in the meridian plane. The color maps have (0.1 $R_J$)$^2$ pixels. Jupiter's disk is displayed in black, the dotted circles around it mark radial distance by 1 $R_J$ steps, the dotted straight lines mark jovicentric latitudes by $10^\circ$ steps, and the bold dotted line is the jovicentric equator, aligned with the jovicentrifugal and jovimagnetic equators in this plane.}
\label{fig:plasma_map}
\end{figure}

\subsection{Synthesizing Occurrence Probability Distributions versus Frequency and Latitude}\label{sec:synth}

The above models allow us to compute the electron density $n_e$, its gradient $\nabla n_e$ and the vector magnetic field $\bf{B}$ at any point of Jupiter's inner magnetosphere. For our study, we computed these quantities over a 3D cartesian grid $-15 \leq x,y \leq +15$ $R_J$ in the equatorial plane and $-10 \leq z \leq +10$ $R_J$ along the rotation axis, with a grid step $dx~=~dy~=~dz~=~0.1$ $R_J$. Figure \ref{fig:plasma_map} displays the meridian distributions (a) of $f_{pe}$ and $n_e$ predicted by Imai's (2016) diffusive density model, (b) of $f_{ce}$ and $B$ distributions by the VIP4 model and (c) of $f_{uh}$ distribution inside the IPT by combining both models. Although it counts 2.3 million points, our grid is coarse because its step is much larger than the wavelength of the nKOM radio waves and cannot capture any detail of the wave-particle or wave-wave interactions involved in their generation.

\begin{figure}[!ht]
\centering
\includegraphics[width=0.75\textwidth]{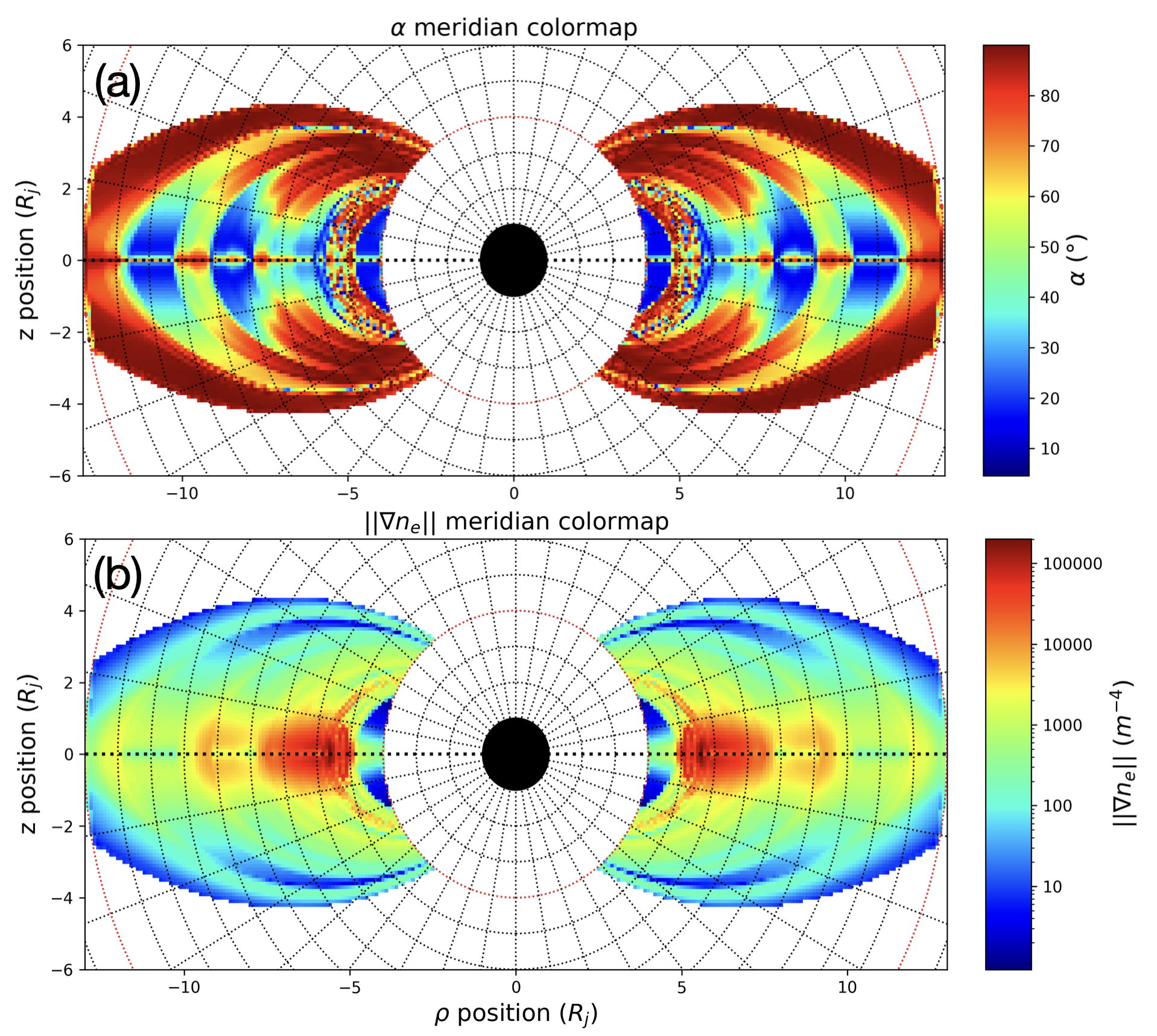}
\caption{Meridian maps of (a) the angle between the magnetic-field and the density gradients $\alpha$, and (b) the density gradient strength $||\nabla n_e||$. The meridian plane, axis, color map parameters and annotations are defined as in Figure \ref{fig:plasma_map}.}
\label{fig:params_map}
\end{figure}

Following the prescriptions of each scenario of Table \ref{tab:cases}, for each Juno/Waves frequency of Figure \ref{fig:ckl}, we identify all points of the grid matching that frequency $\pm6$\%. Juno/Waves frequency channels from 10 kHz to 141 kHz are distributed on a logarithmic scale with a ratio $\times 1.12$ between consecutive channels \cite{kurth2017a}, thus this interval of $\pm6$\% ensures a continuous frequency coverage. These are the potential sources at that frequency. 
We compute the angle $\alpha$ between $\nabla n_e$ and $\bf{B}$, as well as the density gradient strength $||\nabla n_e||$ for each potential source. Figure \ref{fig:params_map} displays the meridian distributions of (a) $\alpha$ and (b) $||\nabla n_e||$. We define $\epsilon$ as the percentile of the $||\nabla n_e||$ distribution at the considered frequency. Figure \ref{fig_appendix:epsilon} illustrates the distribution of $||\nabla n_e||$ versus $f_{pe}$, $f_{uh}$ and $2f_{uh}$, with four particular values of $\epsilon$ indicated as examples. $\alpha$ and $\epsilon$ are the two parameters that will be explored through our modeling. For a selected range of $\alpha$ and $\epsilon$, the corresponding potential sources become active sources. We then launch one or several rays (depending on the scenario, cf. Tables \ref{tab:cases} and \ref{tab_appendix:cases}) from each active source. As ray-tracing would require a much finer grid and be computationally intractable, we propagate these rays in straight line to infinity, unless a ray crosses a grid cell in which the local cutoff frequency $f_{O}~=~f_{pe}$ in O-mode or $f_{X}~=~\sqrt{f_{pe}^2 + (f_{ce}/2)^2} + f_{ce}/2$ in X-mode, is larger than the wave frequency. In that case, the ray is considered absorbed by the plasma.

The directions of the rays are then compared to the jovicentric coordinates of Juno along its 2016-2019 trajectory (sampled by $1.5^\circ$ steps in latitude and $2^\circ$ steps in longitude). Each ray, from each active point source, beamed at an angle of Juno's direction $\leq \delta \theta$ (we used $\delta \theta~=~2^{\circ}$ in our study), as seen from the corresponding point source, is considered to illuminate and thus be detected by the spacecraft if the ray is not absorbed by the plasma before reaching Juno. We count the number of rays thus detected by Juno along its trajectory. This provides us with a modeled time series of nKOM occurrence at the selected frequency. We repeat this operation for each Juno/Waves frequency of Figure \ref{fig:ckl} and obtain a modeled distribution of nKOM occurrence probability versus frequency and latitude. 

The observed distribution may result from the combination of plasma emission generated at different values of $\alpha$ and $\epsilon$. Therefore, for each scenario of Table \ref{tab:cases} (and \ref{tab_appendix:cases}), we perform a parametric study with $0^\circ \leq \alpha \leq 90^\circ$ by steps $\delta \alpha~=~3^\circ$ and 0\% $\leq \epsilon \leq$ 100\% by steps $\delta \epsilon~=~10$\%. Each of the 300 resulting occurrence probability distributions, computed on the same grid as Figure \ref{fig:ckl}, are compared the to observed distribution with a specifically built correlation-inclusion coefficient $C^{*}_{\alpha, \epsilon}$ (see Figure \ref{fig_appendix:corr_inclusion} for a detailed example of the calculation of $C^{*}$). Physically, $C^{*}_{\alpha, \epsilon}$ is justified by the fact that a simulated distribution for a specific selection of parameters $\alpha$ (i.e. angle $\textbf{B}$, $\nabla n_e$) and $\epsilon$ (i.e. $\nabla n_e$ strength) is not expected to coincide with the observed distribution, that may result from plasma emissions generated in a range of values for $\alpha$ and $\epsilon$, but must be included on it. For a single couple ($\alpha$, $\epsilon$), emissions simulated but not observed are not acceptable, whereas emission observed but not simulated are acceptable. Then, an overall occurrence probability distribution is computed by combining the distributions simulated for values of $\alpha$ and $\epsilon$ corresponding to $C^{*}_{\alpha, \epsilon}$ above a given threshold. This overall distribution is then compared with the observed one using the Pearson linear correlation coefficient $C$.

\begin{table}[!ht]
\caption{Basic prescriptions of the studied scenarios.}
\label{tab:cases}
\centering
\begin{tabular}{l l c c}
\hline
Scenario & Reference  & Frequency & Beaming \\
\hline
\#1 & \citeA{jones1980, jones1986, jones1987} & $f_{pe}$  & $\beta~=~(\textbf{k}_+, \textbf{B})$, $\pi~-~\beta~=~(\textbf{k}_-, \textbf{B})$ \\ 
\#2 & \citeA{fung1987}  & $2f_{uh}$  &  $\textbf{k}\perp\textbf{B}$ \\
\#3 & This study & $f_{pe}$  & $\textbf{k}\parallel -\nabla f_{pe}$ \\
\#4 & This study & $f_{uh}$  & $\textbf{k}\parallel -\nabla f_{uh}$ \\
\hline
\end{tabular}
\end{table}

\section{Results}\label{sec:results}

\begin{figure}[!ht]
\centering
\noindent\includegraphics[width=\textwidth]{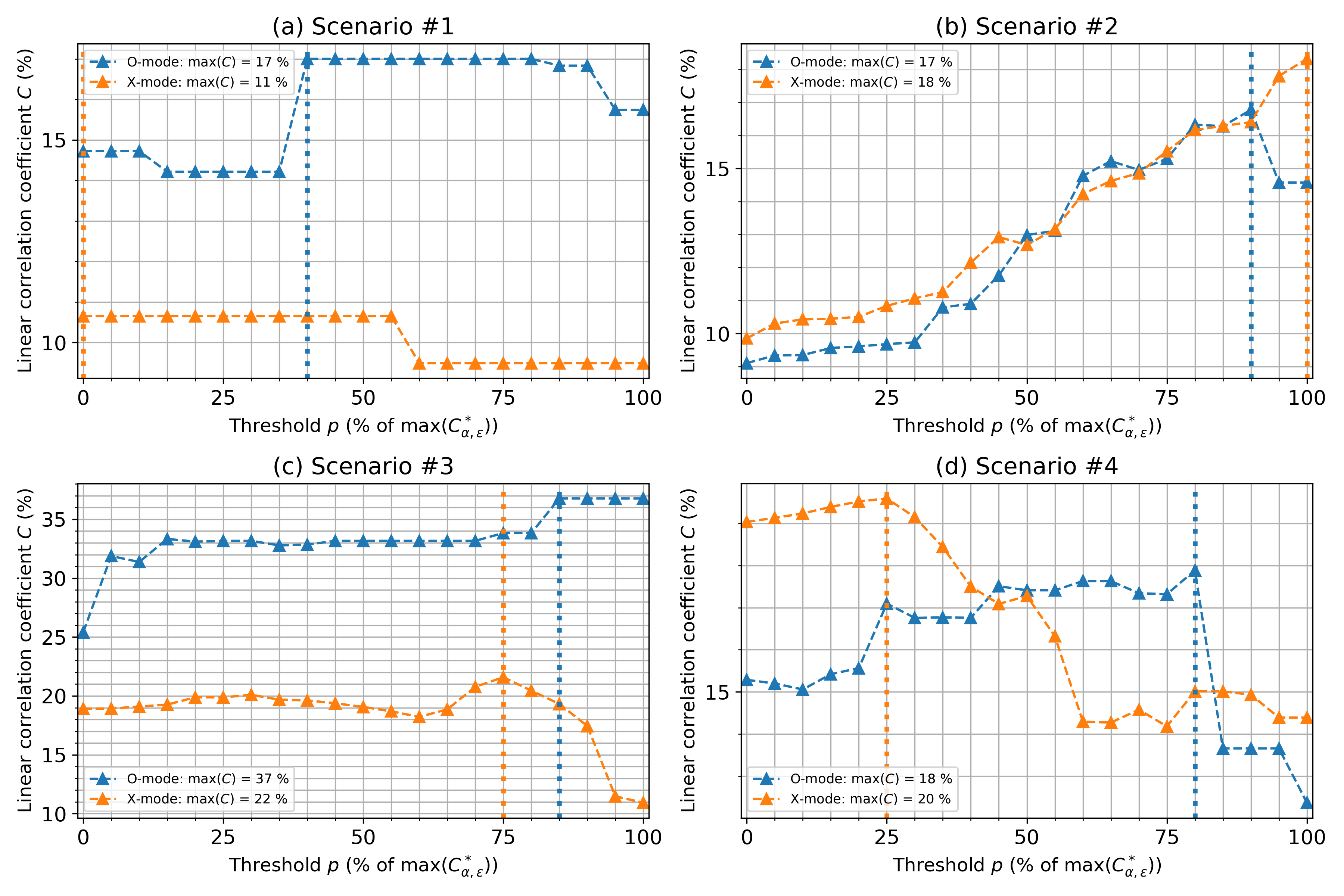}
\caption{Linear correlation coefficient $C$ between the modeled and observed latitude versus frequency distributions of nKOM occurrence, as a function of the threshold $p$ on the correlation-inclusion coefficient $C^{*}_{\alpha, \epsilon}$ defined as a fraction of its maximum value max($C^{*}_{\alpha, \epsilon}$). The blue curves correspond to the cases where we considered the simulated emissions with a cutoff in O-mode and the orange curves in X-mode. Panels (a), (b), (c) and (d) correspond respectively to scenarios 1 to 4.}
\label{fig:corr_plot}
\end{figure}

For each scenario described in Table \ref{tab:cases}, considering the plasma emission cutoff either of O-mode or X-mode, we have combined from the 300 simulated distributions, those for which the value of $C^*_{\alpha, \epsilon}$ is above a threshold $p$. This threshold $p$ is defined as a fraction of max$(C^{*}_{\alpha, \epsilon})$. The evolution of the linear correlation coefficient $C$ with the observations, as a function of $p$, for the overall distribution obtained by selecting ranges of $(\alpha, \epsilon)$ for which $C^*_{\alpha, \epsilon} \ge p$ of max($C^{*}_{\alpha, \epsilon}$), is displayed on the Figure \ref{fig:corr_plot}. The blue curves correspond to the simulations where we considered the emission cutoff in O-mode and the orange curves in X-mode. The dashed vertical lines correspond to the threshold $p$ for which we obtained the highest value of $C$. 

Panel (a) of Figure \ref{fig:corr_plot}, shows that the distributions simulated in the scenario~\#1 are poorly correlated with the nKOM distribution in both O-mode and X-mode. In O-mode $C$ varies slightly with $p$, with a maximum at $C~=~17\%$ for $p=40\%$, and in X-mode $C$ remains roughly constant, with a maximum at $C~=~11\%$ for $p=0\%$. Panel (b) of Figure \ref{fig:corr_plot}, shows that scenario~\#2 distributions are also poorly correlated with the nKOM distribution in both O-mode and X-mode. In both cases, $C$ increases with $p$ until its maximum at $C~=~17\%$ for $p=90\%$ in O-mode, and at $C~=~18\%$ for $p=100\%$ in X-mode. Panel (c) of Figure \ref{fig:corr_plot}, shows that the distributions simulated in the scenario~\#3 are better correlated with the nKOM distribution than for the previous scenarios, especially in O-mode. Indeed, in O-mode, $C$ increases with $p$, with its maximum at $C~=~37\%$ for $p=85\%$, showing a partial correlation with the nKOM distribution. In X-mode, $C$ varies only slightly with $p$ until its maximum at $C~=~22\%$ for $p=75\%$, before decreasing greatly. Finaly, Figure \ref{fig:corr_plot} panel (d), shows that scenario~\#4 distributions are also poorly correlated with the nKOM distribution in both O-mode and X-mode. In O-mode, $C$ increases with $p$ until its maximum value at $C~=~18\%$ for $p=80\%$ and in X-mode, $C$ remains constant until its maximum at $C~=~20\%$ for $p=25\%$, before rapidly decreasing. 

\begin{figure}
\noindent\includegraphics[width=\textwidth]{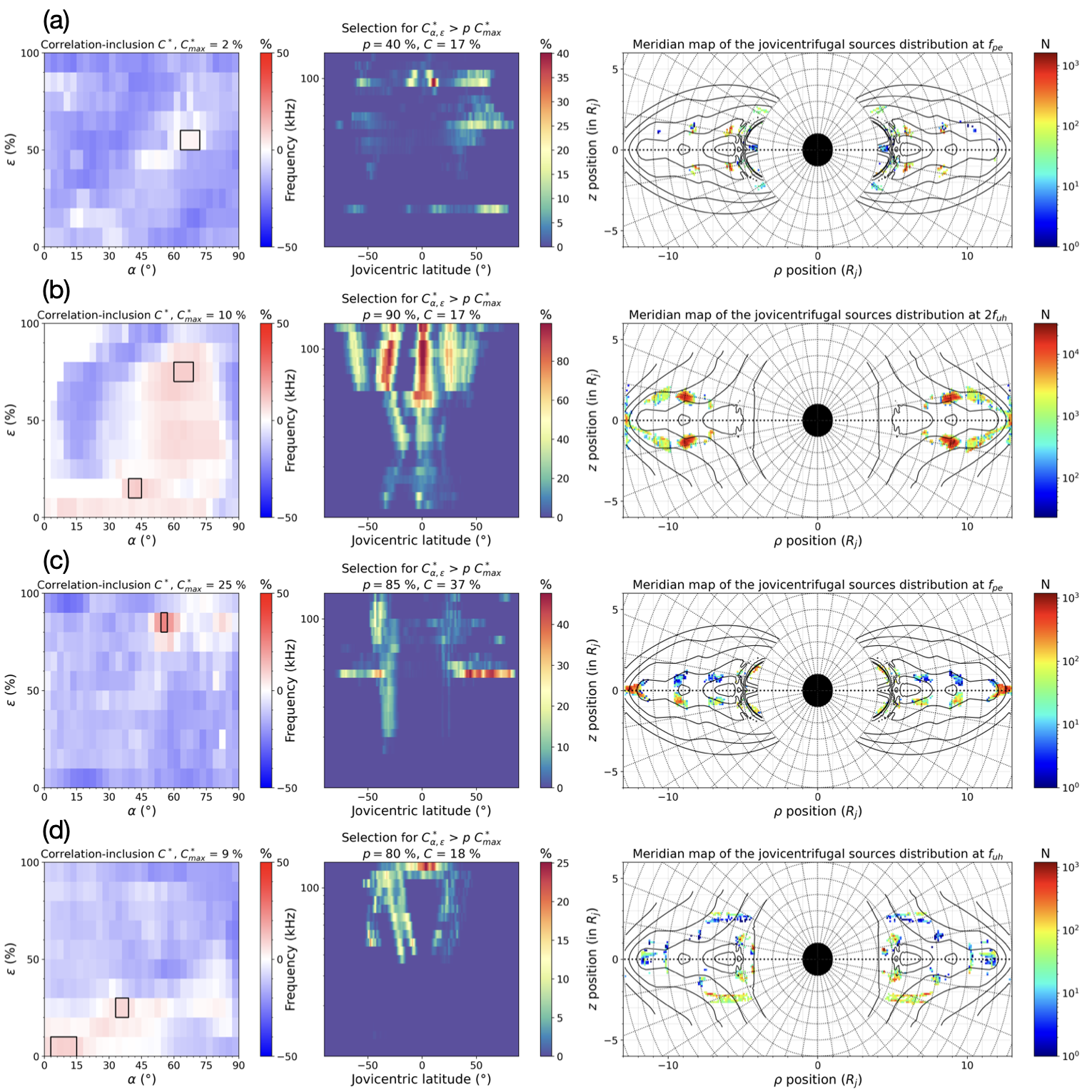}
\caption{O-mode synthetic results derived from Figure \ref{fig:corr_plot}: Rows (a), (b), (c) and (d) correspond respectively to scenarios 1 to 4. (Left column) correlation-inclusion coefficients $C^{*}_{\alpha , \epsilon}$ between the modeled and observed frequency--latitude distributions of nKOM occurrence probability, as a function of $\alpha$ (within $3^{\circ}$ wide bins) and $\epsilon$ (within 10\% wide bins). The regions where $C^{*}_{\alpha , \epsilon} > p$ of max($C^{*}_{\alpha, \epsilon}$), are contoured in black. The value of max($C^{*}_{\alpha, \epsilon}$), annotated $C^*_{max}$, is indicated above each panel of this column. (Central column) simulated distributions of nKOM overall occurrence probability versus frequency and latitude for all values of $\alpha$ and $\epsilon$ within the contoured black region in the left panel considered together. The value of the Pearson correlation coefficient $C$ of this distribution with the observed one of Figure \ref{fig:ckl} is indicated above each panel. The color scale is adjusted in each panel. (Right column) meridian projection color maps of the sources distribution in the IPT in each scenario (indicated above each panel), in the jovicentrifugal coordinates. The black contours correspond the values $f~=~[10, 20, 40, 80, 160, 320]$ kHz of the emitted frequencies in each scenario. In the $2f_{uh}$ case, the contour starts at $f~=~20$ kHz. The color map has (0.1 $R_J$)$^2$ pixels. Jupiter's disk is displayed in black, the dotted circles around it mark radial distance by 1 $R_J$ steps, the dotted straight lines mark jovicentrifugal latitudes by $10^\circ$ steps, and the bold dotted line is the jovicentrifugal equator. The inner IPT extends from $\leq$ 6 to $\sim$ 8 $R_J$, the outer IPT from $\sim$ 8 to $\sim$ 10 $R_J$ and the inner plasma disc beyond 10 $R_J$.}
\label{fig:results_O}
\end{figure}

\begin{figure}[!ht]
\noindent\includegraphics[width=\textwidth]{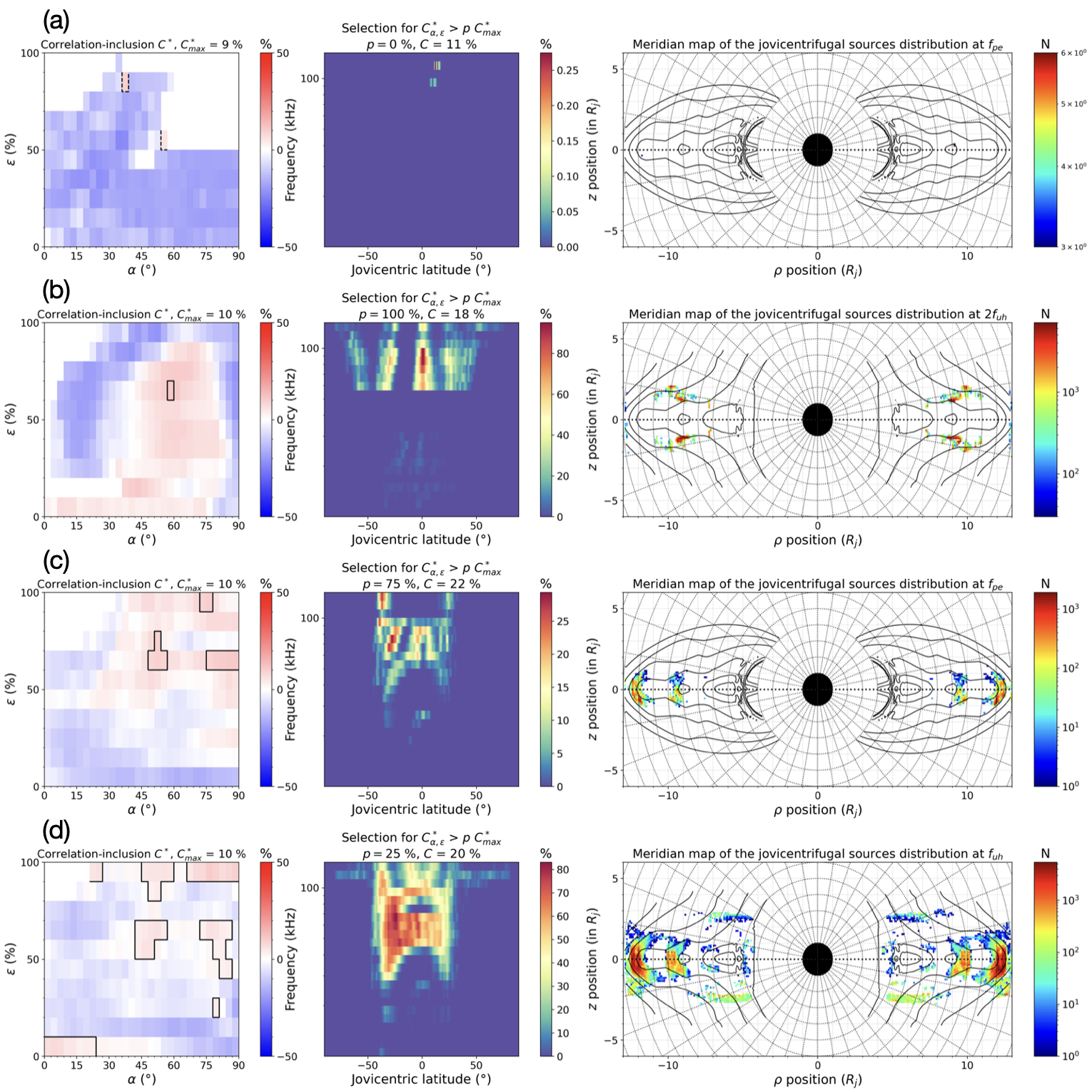}
\caption{Same as Figure \ref{fig:results_O}, but derived from the X-mode modeling results of Figure \ref{fig:corr_plot}.}
\label{fig:results_X}
\end{figure}

Figures \ref{fig:results_O} and \ref{fig:results_X} synthesize the results obtained for each scenario, respectively in O-mode and X-mode, for the thresholds $p$ (cf. Figure \ref{fig:corr_plot}) that return the highest correlation coefficients $C$ of the overall simulated distributions with the observed nKOM. The left column of each figure displays the correlation-inclusion coefficients $C^{*}$ between the modeled and observed frequency--latitude distributions of nKOM occurrence probability, as a function of $\alpha$ (angle between $\nabla n_e$ and $\bf{B}$) and $\epsilon$ (percentile of the density gradient strength $||\nabla n_e||$ distribution at each frequency). On each of these panels, the intervals of $\alpha$ and $\epsilon$ corresponding to the distributions above $p$ are highlighted (with black contours). The central column of each figure displays the simulated distributions of nKOM occurrence probability versus frequency and latitude for the highlighted selections of the left column, with the overall linear correlation coefficient $C$ noted above each panel.
The right side columns display the jovicentrifugal distributions of the sources corresponding to the distributions of the central columns. The black contours correspond to values of the emitted frequencies in each scenario ($f_{pe}$, $f_{uh}$ or $2f_{uh}$) in the meridian plane where the jovicentrifugal and jovimagnetic equators are aligned.
Below, we investigate the 4 scenarios and comment the parameters selection, distributions and source locations obtained in O-mode and X-mode.

\subsection{Scenario~\#1: Jones theory}


The results for scenario~\#1 in O-mode are displayed on row (a) of Figure \ref{fig:results_O}. There is a clear dependence of the correlation-inclusion value on $\alpha$ and $\epsilon$, as the contoured area $C^{*}_{\alpha , \epsilon} > 40\%$ of max($C^{*}_{\alpha, \epsilon}$) (Figure \ref{fig:results_O}a, left) is very localised at moderately large angles $\alpha~=~63^{\circ}~-~72^{\circ}$ and intermediate gradients $\epsilon~=~40\%~-~50\%$. But $C^{*}_{\alpha, \epsilon}$ is mostly negative and the overall distribution (Figure \ref{fig:results_O}a, center), is poorly correlated with the observed nKOM distribution ($C~=~17\%$). Nevertheless, it has some interesting features: it presents a North-South asymmetry, and an occurrence patch $>+30^{\circ}$ in the North at 35--80 kHz. The predicted sources (Figure \ref{fig:results_O}a, right) are distributed from the middle to the inner edge of the IPT. The low frequency sources below 20 kHz are located along the inner edge of the IPT at 4--5 $R_J$, within $\pm 5^{\circ}$ of the jovicentrifugal equator while the high frequency sources above 80 kHz are located at 5--6 $R_J$ and 8--9 $R_J$ for jovicentrifugal latitudes around $\pm10^{\circ}$. The sources responsible for the occurrence patch $>+30^{\circ}$ in the North, are located in the inner edge of the IPT at 4--5 $R_J$ for jovicentrifugal latitudes around $\pm 35^{\circ}$.

The results for scenario~\#1 in X-mode are displayed on row (a) of Figure \ref{fig:results_X}. There seems to be a dependence of $C^{*}_{\alpha, \epsilon}$ on $\alpha$ and $\epsilon$, as the contoured area $C^{*}_{\alpha , \epsilon} > 0\%$ of max($C^{*}_{\alpha, \epsilon}$) (Figure \ref{fig:results_X}a, left), i.e. the regions where the $C^{*}_{\alpha, \epsilon} > 0$, is localised at intermediate angles $\alpha~=~36^{\circ}~-~39^{\circ}$ and $\alpha~=~54^{\circ}~-~57^{\circ}$, and intermediate gradients $\epsilon~=~50\%~-~60\%$ and strong gradients $\epsilon~=~80\%~-~90\%$. However, $C^{*}_{\alpha, \epsilon}$ is mostly negative and the overall distribution (Figure \ref{fig:results_X}a, center) is almost empty, meaning that the majority of the emissions are absorbed by the plasma. The right panel on Figure \ref{fig:results_O}a right, shows that only 12 sources are observed (while the order of magnitude in the other cases is $\sim 1000$). The resulting distribution does not match with the nKOM observations.

\subsection{Scenario~\#2: Fung and Papadopoulos theory}

The results for scenario~\#2 in O-mode are displayed on row (b) of Figure \ref{fig:results_O}. There is a dependence of the correlation-inclusion value on $\alpha$, as the contoured $C^{*}_{\alpha , \epsilon} > 90\%$ of max($C^{*}_{\alpha, \epsilon}$) (Figure \ref{fig:results_O}b, left) correspond to intermediate angles $\alpha~=~39^{\circ}~-~45^{\circ}$ and moderately large angles $\alpha~=~60^{\circ}~-~69^{\circ}$. On the other hand, the dependence of $C^{*}_{\alpha, \epsilon}$ on $\epsilon$ is not convex, as $C^{*}_{\alpha , \epsilon} > 90\%$ of max($C^{*}_{\alpha, \epsilon}$) correspond to both weak gradients $\epsilon~=~10\%~-~20\%$ and strong gradients $\epsilon~=~70\%~-~80\%$. The overall distribution (Figure \ref{fig:results_O}b, center) is poorly correlated with the observed nKOM: it does not present a North-South asymmetry; at low latitudes it presents very strong peaks above 60 kHz and weak peaks down to 15 kHz and it does not predict occurrences at high latitudes below 55 kHz. The predicted sources (Figure \ref{fig:results_O}b, right) are distributed from the middle to the outer edge of the IPT for jovicentrifugal latitudes $< 15^{\circ}$. The sources below 80 kHz are mainly located at the outer edge of the IPT at the limit of validity of Imai's model ($13$ $R_J$). The high frequency sources above 80 kHz are distributed in the middle part of the IPT at 8--10 $R_J$ and 10--12 $R_J$ for jovicentrifugal latitudes within 5$^{\circ}$--13$^{\circ}$.

The results for scenario~\#2 in X-mode are displayed on row (b) of Figure \ref{fig:results_X}. The contoured $C^{*}_{\alpha , \epsilon}~=~\mathrm{max}(C^{*}_{\alpha, \epsilon})$ (Figure \ref{fig:results_X}b, left) correspond to moderately large $\alpha~=~57^{\circ}~-~60^{\circ}$ and strong gradients $\epsilon~=~60\%~-~70\%$. The overall distribution (Figure \ref{fig:results_X}b, center) is poorly correlated with the observed nKOM and shares similar defects with the O-mode case (cf. Figure \ref{fig:results_O}b, center), but with a sharp cutoff in the observation below 50 kHz. The predicted sources (Figure \ref{fig:results_X}b, right) are distributed as in the O-mode case, but with significantly less low frequency sources $>10$ $R_J$.

\subsection{Scenario~\#3: emission at $f_{pe}$ beamed along -$\nabla f_{pe}$}

The results for scenario~\#3 in O-mode are displayed on row (c) of Figure \ref{fig:results_O}. This case clearly stands out, as the correlation-inclusion value seems to be strongly controlled by $\alpha$ and $\epsilon$. The contoured area $C^{*}_{\alpha , \epsilon} > 85\%$ of max($C^{*}_{\alpha, \epsilon}$) (Figure \ref{fig:results_O}c, left) corresponds to $C^{*}_{\alpha , \epsilon}~=~\mathrm{max}(C^{*}_{\alpha, \epsilon})$ localised at intermediate angles $\alpha~=~54^{\circ}~-~57^{\circ}$ and strong gradients $\epsilon~=~80\%~-~90\%$. The overall distribution (Figure \ref{fig:results_O}c, center), shows partial correlation with the nKOM observations as it shares most of its features. It presents a strong North-South asymmetry, an occurrence peak in the South around $-40^{\circ}$ at $20~-~141$ kHz, with a high latitude extension down to $<-70^{\circ}$ at $40~-~50$ kHz, and a very strong peak in the North $>+40^\circ$ at $40~-~50$ kHz. However, some significant features are missing: the distribution does not present any occurrence within $\pm 15^{\circ}$ of the jovicentrifugal equator above 50 kHz and the strong peak in the North does not extend to 30 kHz. The predicted sources (Figure \ref{fig:results_O}c, right) are distributed from the inner edge to the outer edge of the IPT within $\pm 20^{\circ}$ of the jovicentrifugal equator. They can be divided into four groups. The first group is located at the outer edge of the IPT at $11-13$ $R_J$, along the jovicentrifugal equator, and is responsible for the occurrence peak in the South at $-40^{\circ}$ at 20-100 kHz and the weak occurrence above 50 kHz in North near $+30^{\circ}$. The second and third groups are located in the middle part of the IPT, respectively at $9-10$ $R_J$ and $6-8$ $R_J$, for jovicentrifugal latitudes below $\pm 10^{\circ}$. They both contribute to the high occurrence peak above 80 kHz. The fourth group is located at the inner edge of the IPT at $4-6$ $R_J$ below $\pm 40^{\circ}$, following the contour of the low density cavity predicted by Imai's model (cf. Figure \ref{fig:plasma_map}a), and is responsible for the North and South high latitudes occurrence peaks.

The results for scenario~\#3 in X-mode are displayed on row (c) of Figure \ref{fig:results_X}. There is a weak dependence of $C^{*}_{\alpha, \epsilon}$ on $\alpha$ and $\epsilon$, as the contoured area $C^{*}_{\alpha , \epsilon} > 75\%$ of max($C^{*}_{\alpha, \epsilon}$) (Figure \ref{fig:results_X}c, left) corresponds to intermediate angles $\alpha~=~48^{\circ}~-~57^{\circ}$ and large to quasi-perpendicular angles $\alpha~=~72^{\circ}~-~90^{\circ}$, and strong gradients $\epsilon~=~60\%~-~80\%$ and $\epsilon~=~90\%~-~100\%$. The overall distribution (Figure \ref{fig:results_X}c, center), is poorly correlated with the nKOM observations as it does not presents any occurrences at high Northern $>40^{\circ}$ and Southern $<-50^{\circ}$ latitudes. However, some interesting features are present: a peak in the South around $-40^{\circ}$ and in the North around $+15^\circ$ above 50 kHz. Furthermore, it presents low but non-zero occurrences near the equator, that were missing in the O-mode case. The predicted sources (Figure \ref{fig:results_X}c, right) are distributed from the middle to the outer edge of the IPT for jovicentrifugal latitudes below $\pm 5^{\circ}$. The sources below 80 kHz are located at the outer edge of the IPT at 11--13 $R_J$ close to the jovicentrifugal equator. The high frequency sources above 80 kHz are located in the middle part the IPT at 9--10 $R_J$.

\subsection{Scenario~\#4: emission at $f_{uh}$ beamed along -$\nabla f_{uh}$}

The results for scenario~\#4 in O-mode are displayed on row (d) of Figure \ref{fig:results_O}. The contoured $C^{*}_{\alpha , \epsilon} > 80\%$ of max($C^{*}_{\alpha, \epsilon}$) (Figure \ref{fig:results_O}d, left) correspond to quasi-parallel angles $\alpha~=~3^{\circ}~-~15^{\circ}$ and intermediate angles $\alpha~=~33^{\circ}~-~39^{\circ}$, and weak gradients $\epsilon~=~0\%~-~10\%$ and $\epsilon~=~20\%~-~30\%$. The overall distribution (Figure \ref{fig:results_O}d, center) is poorly correlated with the observed nKOM: it does not present an inverted North-South asymmetry, it presents a very strong peak around 140 kHz near the equator and it does not predict occurrences at high latitudes below 55 kHz. The sources (Figure \ref{fig:results_O}d, right) are scattered from the inner edge to the outer edge of the IPT at very different jovicentrifugal latitudes. 

The results for scenario~\#4 in X-mode are displayed on row (d) of Figure \ref{fig:results_X}. The dependence of $C^{*}_{\alpha, \epsilon}$ on both $\alpha$ and $\epsilon$ is noisy, as the contoured area $C^{*}_{\alpha , \epsilon} > 75\%$ of max($C^{*}_{\alpha, \epsilon}$) (Figure \ref{fig:results_X}d, left) corresponds to values of $\alpha$ and $\epsilon$ spread all over the parameter space. The overall distribution (Figure \ref{fig:results_X}d, center), is poorly correlated with the nKOM observations as it resembles to a smoother version the distribution obtained for the scenario~\#3 in X-mode (Figure \ref{fig:results_X}c, center). The predicted sources (Figure \ref{fig:results_X}d, right) are distributed from the inner edge to the outer edge of the IPT at very different jovicentrifugal latitudes in a similar way as for the scenario~\#3 in X-mode (Figure \ref{fig:results_X}c, right). The sources below 90 kHz are located at the outer edge of the IPT at 11--13 $R_J$ close to the jovicentrifugal equator. The high frequency sources above 90 kHz are located in the middle part the IPT at 9--10 $R_J$. In addition, some sources in 60--140 kHz are distributed from the middle to the inner part of the IPT in 5--8 $R_J$ at jovicentrifugal latitudes around $\pm 25^{\circ}$.

\section{Discussion and Perspectives}

We have developed a 3D numerical large-scale model to test the nKOM generation mechanism and beaming, based on the plasma density model of \citeA{imai2016} and the VIP4 internal Jovian magnetic field model \cite{connerney1998} (cf. Figure \ref{fig:plasma_map}). We have tested the model predictions against the observed frequency--latitude nKOM occurrence probability measured by Juno/Waves over the first 3 years of its Jovian tour (cf. Figure \ref{fig:ckl}). We have explored 4 emission scenarios based on assumed emission frequencies and beaming patterns (cf. Table \ref{tab:cases}), for each of them we have either considered the emission cutoff in O-mode or X-mode, and performed a parametric study versus the angle $\alpha$ between $\bf{B}$ and $\nabla n_e$, and the (large-scale) density gradient strength percentile $\epsilon$ at each frequency (cf. Figure \ref{fig_appendix:epsilon}). 

Our results allow us to exclude scenario~\#1, based on the Jones theory of linear mode conversion of Z-mode electrostatic waves at the upper-hybrid resonance, to O-mode electromagnetic waves via the ``radio window" \cite{jones1980,jones1986,jones1987}. This theory postulates very specific beaming angles ($\textbf{k}, \textbf{B}$) that, in O-mode, are not supported by the comparison of all modeled occurrence probability distributions with the observed one, especially for the emissions observed at low latitudes. It predicts a high Northern latitude occurrence peak that resembles the nKOM observations with slightly higher frequency, but it requires large but not perpendicular angles $\alpha$ and relatively strong gradient $\epsilon$, that are incompatible with this theory. In X-mode, all the distributions modeled are either empty or anti-correlated with the nKOM, confirming that this theory cannot be used to describe the beaming for conversion from Z-mode to X-mode.

We can also confidently exclude scenario~\#2, based on \citeA{fung1987} theory that predicts emission at $2 f_{uh}$ (resulting from nonlinear coupling of two electrostatic waves at $f_{uh}$) beamed in a plane nearly perpendicular to the local $\bf{B}$. The modeled occurrence probability distributions, in both O-mode and X-mode, do not match the observed one, primarily because emissions at $2 f_{uh}$ cover a too broad frequency range compared to nKOM emissions, and secondly because its beaming does not predict nKOM low-frequency extensions at high latitudes. We can extrapolate this result to exclude emissions at $2 f_{pe}$ or any higher harmonic of $f_{pe}$ or $f_{uh}$, to be nKOM below 141 kHz.

Finally, we exclude scenario~\#4, i.e. plasma emissions at $f_{uh}$ beamed along their local negative frequency gradient (i.e. toward decreasing frequencies). The modeled distribution in O-mode is not compatible with the nKOM observations at high and low latitudes. The modeled distributions in X-mode, on the other hand, bear some resemblance with the nKOM distribution at mid and low latitudes from $-50^{\circ}$ to $<+30^{\circ}$. However, the distribution of $\alpha$ and $\epsilon$ leading to the best simulated distribution are scatter, which is difficult to justify physically.

Only the scenario~\#3, i.e. plasma emissions at $f_{pe}$ beamed along their local negative frequency gradient, clearly stands out, being the only case to show a significant positive correlation with the nKOM observations (with $C~=~37\%$ in O-mode). It favours O-mode emission to describe the nKOM occurrences peak observed at high latitudes, from regions of relatively strong electron density gradients ($\epsilon~=~80\%~-~90\%$) at an angle $\alpha \sim 55^\circ$ from the local magnetic field. The angle $\alpha$ is lower than the $90^\circ$ value assumed in the Jones theory, as critical coupling at $f~=~f_{pe}$ requires that $\sin(\alpha) \gg \cos(\alpha)$. These nKOM sources are predicted to be distributed at the inner edge of the IPT ($<~6~R_J$), within $\pm 40^{\circ}$ of the jovicentrifgual equator. The X-mode modeled distribution, although poorly correlated with the observed one, displays nKOM at mid-to-low latitudes from $-50^{\circ}$ to $+30^{\circ}$.

\begin{figure}[!ht]
\centering
\noindent\includegraphics[width=0.80\textwidth]{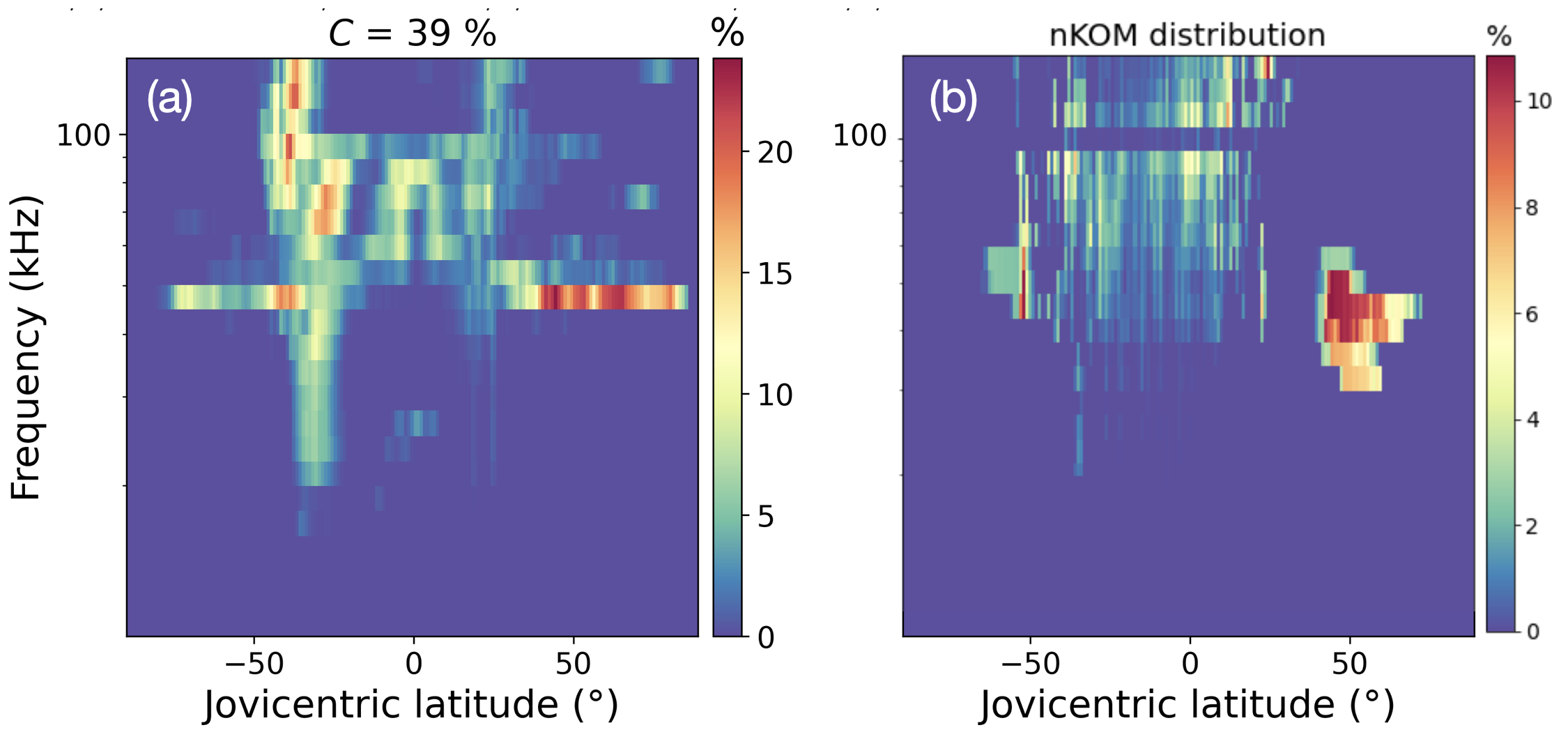}
\caption{(a) Scenario~\#3 combining O-mode and X-mode occurrence probability distribution versus frequency and latitude, respectively Figures \ref{fig:results_O}c and \ref{fig:results_O}d center panels. The nKOM occurrence probability distribution versus frequency and latitude from Figure \ref{fig:ckl}, is displayed for comparison (b).}
\label{fig:scenario3_O_X}
\end{figure}

In the scenario~\#3, the overall O-mode and X-mode distributions seems complementary: the O-mode distribution (cf. Figure \ref{fig:results_O}c, center) is missing the low latitudes emissions, and the X-mode distribution (cf. Figure \ref{fig:results_X}c, center) is missing the low frequency high latitudes emissions. The combined distributions, displayed on Figure \ref{fig:scenario3_O_X} (with the nKOM distribution for comparison), reaches a higher correlation with $C~=~39\%$, and possesses all the features observed on the nKOM distribution. This simulated distribution is not attainable with O-mode emissions alone, as generations parameters that permits the observations of high frequency O-mode at low latitudes, also activate sources that are observed at high latitudes but inconsistent with the nKOM observations (i.e. the simulated distributions have negative $C^{*}_{\alpha, \epsilon}$ values). However, as these high latitudes emissions are beamed from regions where $f_X > f_{pe}$, their observation are restricted by the X-mode cutoff frequency. This suggests that nKOM is produced at $f_{pe}$, in both O-mode and X-mode, with the O-mode sources distributed from the outer edge to the inner edge of the IPT, and the X-modes sources located at the outer edge of the IPT near the jovicentrifugal equator. These results are compatible with the updated \citeA{budden1986} radio window theory, suggesting Z-mode conversion into both O-mode and X-mode, if the density gradients are large. It also confirms and precises the source locations qualitatively proposed by \citeA{kaiser1980} and \citeA{daigne1986}, and also assumed by \citeA{jones1980}, in the middle part of the IPT at 8--9 $R_J$ for O-mode nKOM near 100 kHz, and the goniopolarimetric results of \citeA{reiner1993} within the middle to the outer part of the IPT 7--13 $R_J$ for X-mode nKOM.

By construction, our model cannot capture the entire physics in the sources. Moreover, we assumed straight line radio propagation, which is validated \textit{a posteriori} by the fact that the waves are mostly emitted away from the centrifugal equator, towards regions of lower density. In spite of these limitations, our large-scale 3D modeling of the conditions of emission of nKOM allowed us to discriminate between different models, exclude the two main models proposed so far for nKOM generation, and to reach a reasonable match of the modeled and observed distributions of nKOM occurrence probability with simple assumptions. 
The fact that we obtain modeled distributions compatible with the observed one under the assumption of radio propagation along $- \nabla f_{pe}$ at the source can be due, in the context of our large (0.1 $R_J$) grid cells, to a refraction effect close to the source. Snell-Descartes law shows that waves beaming tends to align with the normal to the iso-surfaces of the refractive index $N$ when $\nabla N$ is large. When applied to a wave in perpendicular propagation in a cold collisionless magnetized plasma, this condition is met at the conversion layer when $f_{pe} \ge f_{ce}$ for O-mode and X-mode, if $f \sim f_{X}$ for X-mode. The refractive index $N$ varies in the same direction as $f_{pe}$ for O-mode ($N^2_O~=~1-f_{pe}^2/f^2$), and for X-mode but only if $f \gg f_{ce}$ (in which case $N^2_O \sim N^2_X$), which is the case for emissions at $f_{pe}$ from the outer edge to the middle part of the IPT, near the jovicentrifugal equator (see Figure \ref{fig_appendix:plasma_map}). If the plasma density strongly decreases locally to the conversion layer, the refractive index reaches $N~=~1$ quickly, and from there, because the emission is considered propagating in vacuum medium, the emission will be observed exiting the plasma with a beaming close to $-\nabla f_{pe}$. As there is no obvious dependence of $N_O$ and $N_X$ with $f_{uh}$, this would explain why the results of scenario~\#4 do not fit the nKOM observations. Finally, this explanation is not valid for waves in parallel propagation unless $f_{ce} \ll f_{pe}$, suggesting that the O-mode nKOM in produced at the inner edge of the IPT necessarily has a perpendicular propagation with the magnetic field locally at the radio sources. This remains to be tested in future studies via small-scale wave propagation close to the sources, for different initial beaming angles, with a ray-tracing code taking into account the refraction of the waves, like ARTEMIS-P \cite{gautier2013} or with measurement inside the sources.

\begin{figure}[!ht]
\noindent\includegraphics[width=\textwidth]{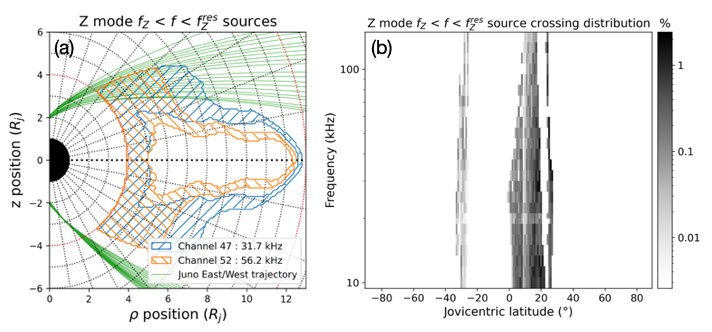}
\caption{(a) Meridian projection of the Juno trajectory in green, crossing the Z-mode sources of interest  at $f~=~31.7$ kHz in orange, and at $f=56.2$ kHz in blue, with (b) occurrence probability versus frequency and latitude of the Z-mode emissions when Juno is crossing the IPT. The panel (a) meridian plane here is where the jovicentric, jovicentrifugal and jovimagnetic equator are aligned (defined by the system III longitude $\varphi~=~200.769^{\circ}$). The Z-mode sources are predicted using the $f_{pe}$ and $f_{ce}$ distributions predicted by \citeA{imai2016} and VIP4 \cite{connerney1998} models (cf. panels (a) and (b) on Figure \ref{fig:plasma_map}). The Z-mode frequencies are defined by the Juno/Waves channels (cf. Sec. \ref{sec:synth}), considering $f_Z<f<f^{res}_Z$ with $f_Z~=~\frac{f_{ce}}{2}[(1+\frac{4 f^2_{pe}}{f^2_{ce}})^{1/2}~-~1]$ and $f^{res}_Z~=~\frac{f_{ce}}{\sqrt{2}}\{(\frac{f^2_{uh}}{f^2_{ce}} + [(1~-~\frac{f^2_{pe}}{f^2_{ce}})^2~-~\frac{4 f^2_{pe}}{f^2_{ce}}\sin^2\theta]\}$. In panel (b) each occurrence correspond to a point of the Juno trajectory inside the plasma which we considered the Z-mode in $f_Z<f<f^{res}_Z$.}
\label{fig:juno_Z_cross}
\end{figure}

Finally, we supposed nKOM to be either O-mode or X-mode emissions, as Z mode waves are trapped in IPT and thus cannot be observed from outside. However, Juno crossed the IPT multiple times when close to Jupiter. In these regions, Juno might have observed Z mode emissions when crossing their sources. Based on VIP4 and Imai's density model, we can estimate the characteristic frequencies of the Z-mode radio sources crossed by Juno in the IPT. In the considered time range (from 9 April 2016 to 24 June 2019), Figure \ref{fig:juno_Z_cross} displays on panel (a) the Juno trajectory when crossing Z-mode source, and on panel (b) the Z-mode occurrence probability versus frequency and latitude when Juno is crossing the IPT. Figure \ref{fig:juno_Z_cross} panel (b), shows that Z-mode emission should mainly be observed in the North around $+20^{\circ}$ in 10-141 kHz. However, despite considering the Z-mode radio sources permanent, the resulting distribution have low probability of occurrences that does not exceed 1\%, which is low in comparison the nKOM observed at the same latitudes (cf. Figure \ref{fig:ckl}). Then, it is a valid assumption to consider the nKOM observed by Juno being either O-mode or X-mode.

It will also be interesting to check the robustness of our results by reproducing the present study with alternate plasma density models (such as \citeA{garrett2016,jun2019} or \citeA{huscher2021}), and magnetic field models (such as JRM09 \cite{connerney2018} or JRM33 \cite{connerney2022}). Alternate plasma models might allow us to model the electron distribution over a radial range broader than 4--13 $R_J$, including the plasmasphere and cool IPT below $\sim5.5$ $R_J$, and a larger part of the inner plasma disc (e.g., up to 20 $R_J$). 
Such an extension could be relevant for addressing the conditions of emission of the narrowband low-frequency emissions discovered by Juno, refereed to as ECR by \citeA{imai2017} and nLF by \citeA{louis2021}. These narrowband emissions have a frequency-latitude distribution of their occurrence probability close to that of nKOM \cite{louis2021}, but have lower frequencies (kHz to tens of kHz). We suspect that they are also plasma emissions, and a study similar to the present one will be dedicated to nLF generation.
Extending the observed distribution of Figure \ref{fig:ckl} (and of nLF) with four more years of observations (i.e. over the interval 2016-2023) will also improve the observational constraints.
Furthermore, before the end of its extended orbital tour (in 2025), Juno should cross the regions where we expect to find the sources of nKOM emission. If nKOM sources are crossed, it will be possible to study \textit{in situ} their local characteristics (frequency, beaming, electrostatic waves, ambient plasma) and precise their emission mechanism. 
Finally, a similar study could be performed at Saturn, using a plasma model such as \citeA{persoon2020}, for constraining the emission conditions of its narrowband low-frequency radio components \cite{wu2021}.

\appendix

\vspace{10cm}
\section{Detailed Table of the Scenario Beaming Characteristics}

\begin{table}[!htbp]
\caption{Detailed version of Table \ref{tab:cases}, expliciting the beaming characteristics for each studied scenario. Scenario~\#1 is based on the theory of \citeA{jones1987} that predicts Jovian plasma emissions produced at $f_{pe}$ and beamed in two opposite directions by making an angle $\beta~=~\arctan (f_{pe}/f_{ce})^{1/2}$ and $\pi~-~\beta$ with respect to $\bf{B}$ into the plane characterized by ($\nabla n_e$,$\textbf{B}$). Scenario~\#2 is based on the theory of \citeA{fung1987} that predicts Jovian plasma emissions produced at $2f_{uh}$ and beamed in a plane perpendicularly to $\bf{B}$;  
for our simulations we assumed that the emissions are beamed in 8 directions in the plane perpendicular to $\bf{B}$. Scenarios \#3 and \#4 predict plasma emissions produced respectively at $f_{pe}$ and $f_{uh}$, and beamed along their local frequency gradient independently from $\bf{B}$ ($\bf{B}$ is represented on the sketches \#3 and \#4 for an illustrative purpose only, but its orientation can be arbitrary).}
\centering
\begin{tabular}{l c c c}
\hline
Scenario : Reference & Beams/Source & Beaming & Scheme  \\
\hline \\
\#1 : \citeA{jones1980, jones1986, jones1987} & 2  & \raisebox{-.5\height}{\shortstack{$\beta~=~(\textbf{k}_+, \textbf{B})$\\ $\pi~-~\beta~=~(\textbf{k}_-, \textbf{B})$}} & \raisebox{-.5\height}{\includegraphics[width=1.90cm]{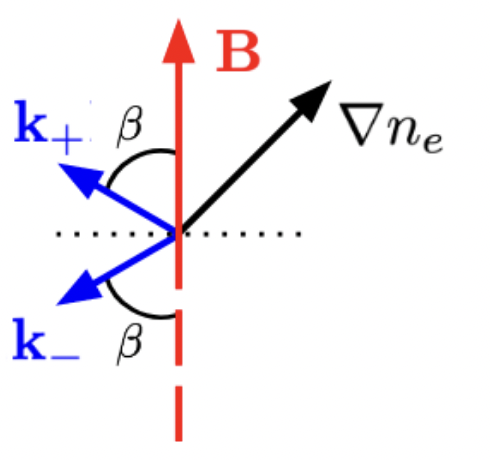}} \\ 
\#2 : \citeA{fung1987}  & 8  &  $\textbf{k}\perp\textbf{B}$ & \raisebox{-.5\height}{\includegraphics[width=1.90cm]{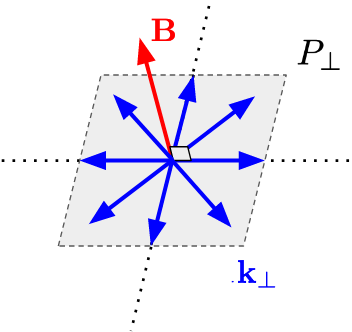}}\\
\#3 : This study & 1  & $\textbf{k}\parallel~-~\nabla f_{pe}$ & \raisebox{-.5\height}{\includegraphics[width=1.90cm]{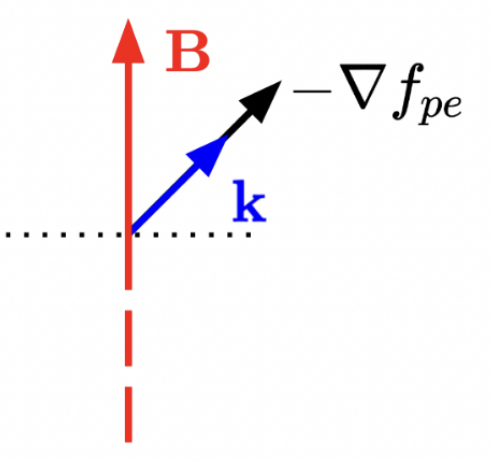}}\\
\#4 : This study & 1  & $\textbf{k}\parallel~-~\nabla f_{uh}$ & \raisebox{-.5\height}{\includegraphics[width=1.90cm]{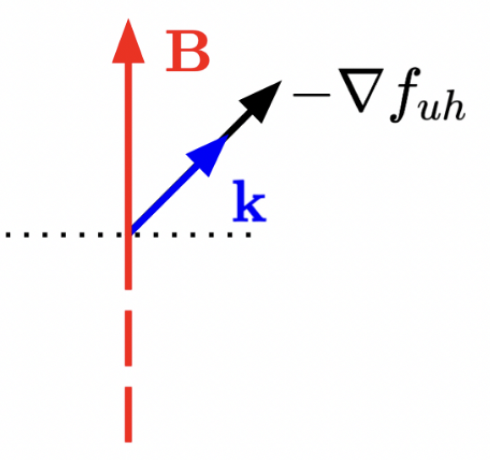}}\\
\hline
\label{tab_appendix:cases}
\end{tabular}
\end{table}

\vspace{10cm}
\section{Density Gradient Strength Distributions}

\begin{figure}[!htbp]
\centering
\noindent\includegraphics[width=\textwidth]{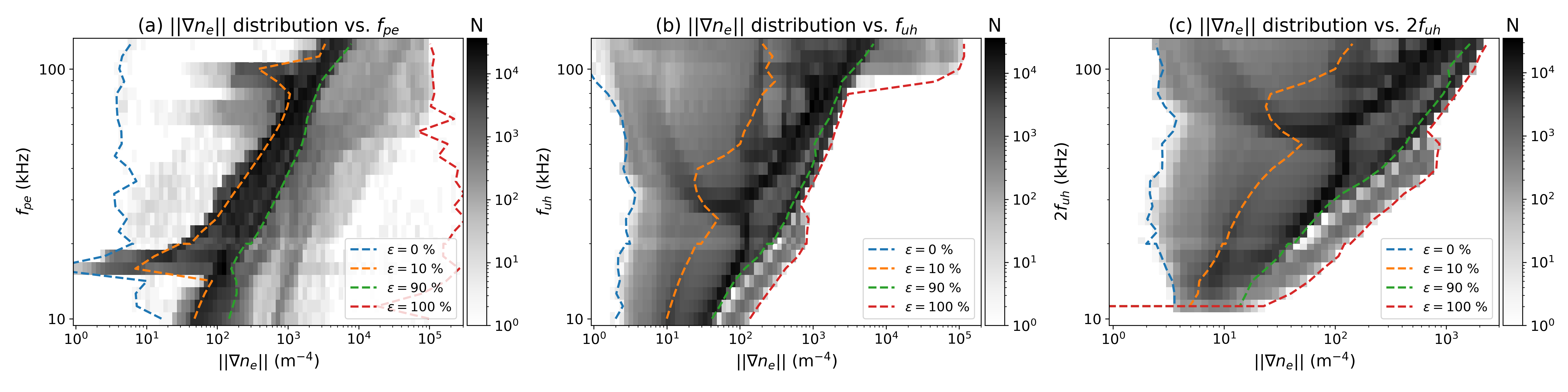}
\caption{Distributions of the number of grid cells versus the density gradient strength $||\nabla n_e||$ and (a) the local plasma frequency $f_{pe}$, (b) the local upper-hybrid frequency $f_{uh}$ and (c) the second harmonic of the upper-hybrid frequency $2f_{uh}$, inside our simulation domain. The horizontal axis correspond to $||\nabla n_e||$ values distributed on a logarithmic scale with a ratio $\times 1.16$ between consecutive bins. The vertical axis correspond to Juno/Waves frequencies from 10 kHz to 141 kHz, distributed on a logarithmic scale with a ratio $\times 1.12$ between consecutive channels.
The color scale indicates the number of grid cells $N$ at each frequency, for each value of $||\nabla n_e||$.
The dashed lines annotated $\epsilon~=~0\%$ in blue, $\epsilon~=~10\%$ in orange, $\epsilon~=~90\%$ in green and $\epsilon~=~100\%$ in red, show respectively as examples, the $0^{th}$, $10^{th}$, $90^{th}$ and $100^{th}$ percentiles of the $||\nabla n_e||$ distribution at each frequency.}
\label{fig_appendix:epsilon}
\end{figure}

\section{Correlation-Inclusion Coefficient}
\begin{figure}[!htbp]
\centering
\noindent\includegraphics[width=0.40\textwidth]{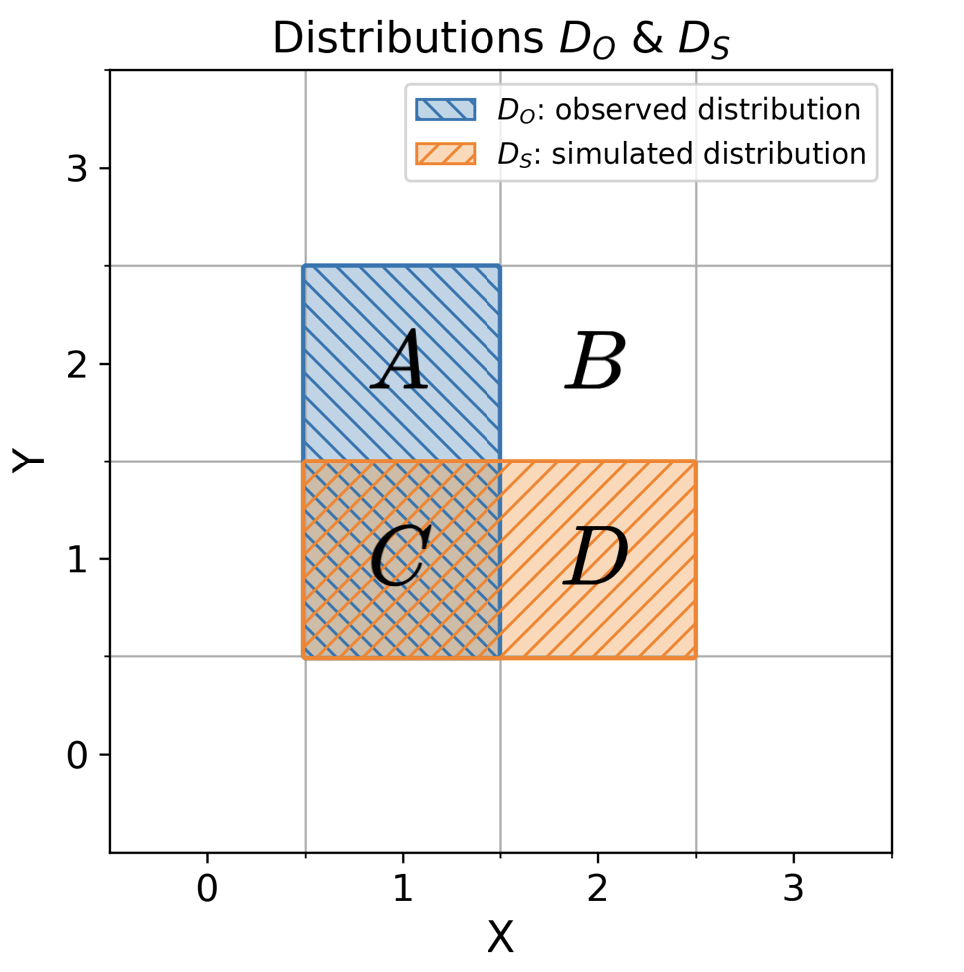}
\caption{Illustration of the principle of the correlation-inclusion coefficient $C^*$ between a modeled $D_S$ and observed distribution $D_O$. $AC$ represents the region where the observed distribution $D_O~=~1$ (equal to 0 elsewhere). $CD$ represents the region where the simulated distribution $D_S~=~1$ (equal to 0 elsewhere). $C^*$ corresponds to the Pearson correlation coefficient $C$ computed after weighting negatively the regions where $D_S \neq 0$ and $D_O~=~0$ (i.e. emissions predicted by the simulation but not observed, here corresponding to the region $D$), meaning that $D_O$ is set negative in this region. Following empirical tests, we have set this values to $-\sigma_O$. In the displayed example, $C^*~=~25\%$ while $C \simeq 43 \%$.}
\label{fig_appendix:corr_inclusion}
\end{figure}

\section{Plasma and Cyclotron Frequency Ratio and X-mode Frequency Cutoff Meridian Maps}
\begin{figure}[!htbp]
\centering
\noindent\includegraphics[width=0.75\textwidth]{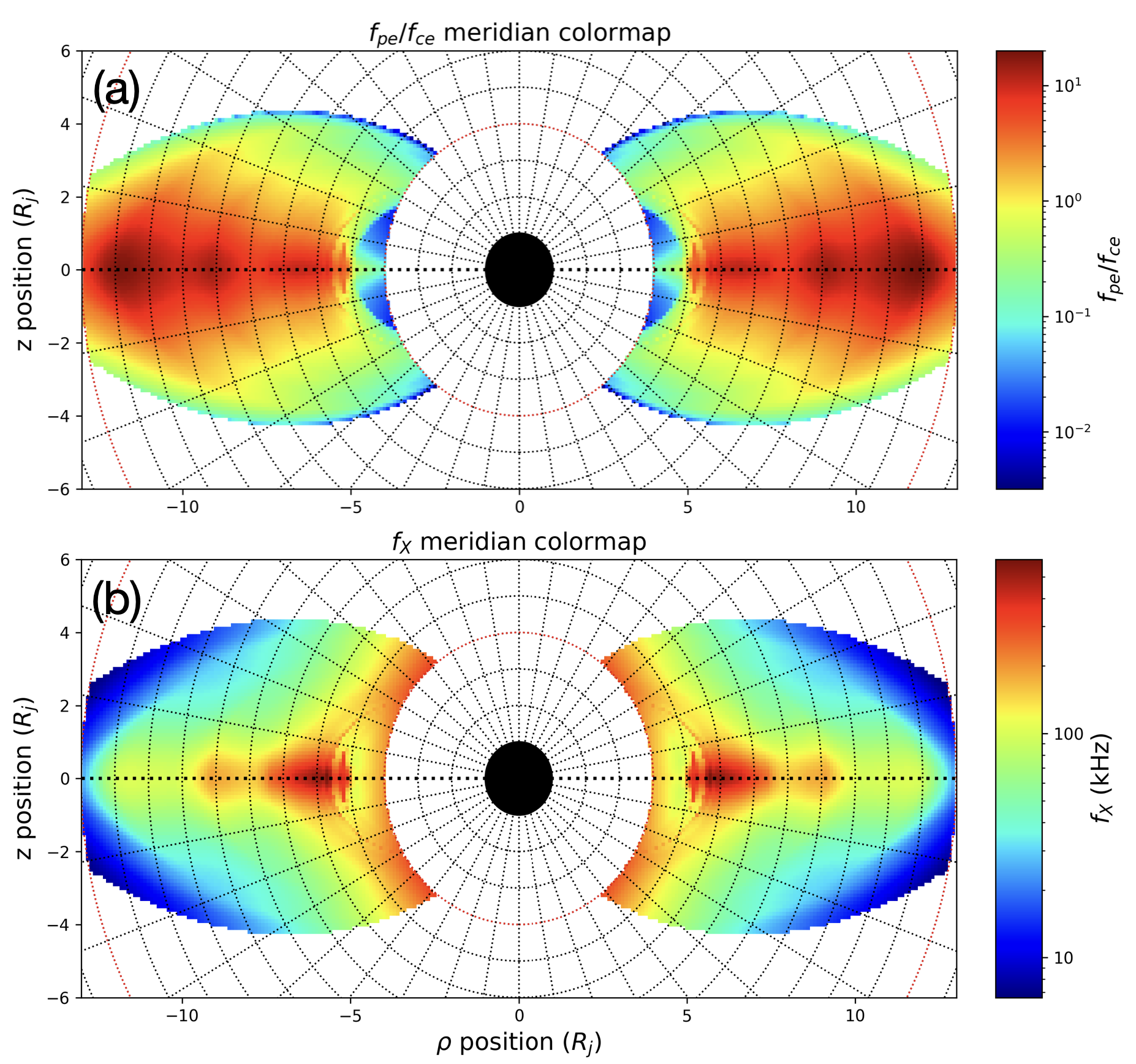}
\caption{Meridian maps of (a) the ratio between the electron plasma frequency $f_{pe}$ and the electron cyclotron frequency $f_{ce}$, (b) the X-mode cutoff frequency $f_{X}~=~\sqrt{f_{pe}^2 + (f_{ce}/2)^2} + f_{ce}/2$. The meridian plane here is where the jovicentric, jovicentrifugal and jovimagnetic equators are aligned (defined by the system III longitude $\varphi~=~200.769^{\circ}$). The vertical axis is the cartesian coordinate $z$. The horizontal axis is the radial cylindrical coordinate $\rho$ in the meridian plane. The color maps have (0.1 $R_J$)$^2$ pixels. Jupiter's disk is displayed in black, the dotted circles around it mark radial distance by 1 $R_J$ steps, the dotted straight lines mark jovicentric latitudes by $10^\circ$ steps, and the bold dotted line is the jovicentric equator, aligned with the jovicentrifugal and jovimagnetic equators in this plane.}
\label{fig_appendix:plasma_map}
\end{figure}

%
%

%

%


\section*{Data Availability Statement}

The Juno/Waves calibrated data used in the manuscript is available at \url{https://doi.org/10.25935/6jg4-mk86} \cite{louis2021_data}. The Juno/Waves catalog used in the manuscript is available at \url{https://doi.org/10.25935/6jg4-mk86} \cite{louis2021_catalog}.

\acknowledgments

We thank L. Lamy, R. Prangé, C. Jackman and her team at DIAS, and the DIAS-LESIA-INPE ``Radioclub" for fruitful and friendly discussions at various steps of this study. We also thank D. Santos Costa and H. Garrett for their insights on Jovian plasma density models. C.K. Louis's work at the Dublin Institute for Advanced Studies was funded by Science Foundation Ireland Grant 18/FRL/6199. \add[Imai]{The work of M. Imai was supported by the JSPS KAKENHI Grant Number JP23K13164.}


%
%



\bibliography{agusample.bib}

%
%
%
%
%

\end{document}


%
%


\title{Supporting Information for "Generation mechanism and beaming of Jovian nKOM from 3D numerical modeling of Juno/Waves observations"}
%
%

%
%



\authors{A. Boudouma\affil{1}, P. Zarka\affil{1,2}, C.K. Louis\affil{3}, C. Briand\affil{1}, M. Imai\affil{4}}


\affiliation{1}{LESIA, Observatoire de Paris, CNRS, PSL, Sorbonne Université, Université Paris Cité, Meudon, France}
\affiliation{2}{Observatoire Radioastronomique de Nançay, ORN, Observatoire de Paris, CNRS, PSL, Université d'Orléans, Nançay, France}
\affiliation{3}{School of Cosmic Physics, DIAS Dunsink Observatory, Dublin Institute for Advanced Studies, Dublin, Ireland}
\affiliation{4}{Dept of Electrical Engineering and Information Science, National Institute of Technology (KOSEN), Niihama College, Niihama, Japan}

%
%

%

\begin{article}

%
%

\noindent\textbf{Contents of this file}
\begin{enumerate}
\item Table S1 : Expanded Table 1 with details on the beaming characteristics for each scenario.
\item Figure S1 : Simple example to illustrate the use of the correlation-inclusion coefficient to compare the simulated distributions with the observations.
\item Figure S2 : Density gradient strength occurrence distributions versus the local plasma frequency, the upper-hybrid frequency and the second harmonic of the upper-hybrid frequency.
\item Figure S3 : Meridian color map of the ratio between the local plasma frequency and cyclotron frequency, and the extraordinary mode cutoff frequency.

\end{enumerate}


%
%


%
%
\bibliography{agusample}
%
%
%


%
%
%
%
%

%
%
\end{article}

\begin{table}
\settablenum{S1} 
\caption{Detailed version of Table 1, expliciting the beaming characteristics for each studied scenario. Scenario \#1 is based on the theory of \citeA{jones1987} that predicts Jovian plasma emissions produced at $f_{pe}$ and beamed in two opposite directions by making an angle $\beta = \arctan(\sqrt{(f_{pe}/f_{ce})^2})$ and $\pi - \beta$ with respect to $\bf{B}$ into the plane characterized by ($\nabla n_e$,$\textbf{B}$). Scenario \#2 is based on the theory of \citeA{fung1987} that predicts Jovian plasma emissions produced at $2f_{uh}$ and beamed in a plane perpendicularly to $\bf{B}$ ;  
for our simulations we assumed that the emissions are beamed in 8 directions in the plane perpendicular to $\bf{B}$. Scenarios \#3 and \#4 predict plasma emissions produced respectively at $f_{pe}$ and $f_{uh}$, and beamed along their local frequency gradient independently from $\bf{B}$ ($\bf{B}$ is represented on the sketches \#3 and \#4 for an illustrative purpose only, but its orientation can be arbitrary).}
\centering

\begin{tabular}{l c c c}
\hline
Scenario : Reference & Beams/Source & Beaming & Scheme  \\
\hline \\
\#1 : Jones 1980, 1986, 1987 & 2  & \raisebox{-.5\height}{\shortstack{$\beta = (\textbf{k}_+, \textbf{B})$\\ $\pi - \beta = (\textbf{k}_-, \textbf{B})$}} & \raisebox{-.5\height}{\includegraphics[width=3cm]{figures/annexe/scenario1.png}} \\ 
\#2 : Fung \& Papadopoulos 1987  & 8  &  $\textbf{k}\perp\textbf{B}$ & \raisebox{-.5\height}{\includegraphics[width=3cm]{figures/annexe/scenario2.png}}\\
\#3 : -- & 1  & $\textbf{k}\parallel - \nabla f_{pe}$ & \raisebox{-.5\height}{\includegraphics[width=3cm]{figures/annexe/scenario3.png}}\\
\#4 : -- & 1  & $\textbf{k}\parallel - \nabla f_{uh}$ & \raisebox{-.5\height}{\includegraphics[width=3cm]{figures/annexe/scenario4.png}}\\
\hline
\label{tab_si:cases}
\end{tabular}
\end{table}

\begin{figure}
\centering
\noindent\includegraphics[width=0.5\textwidth]{figures/annexe/annoted_distribution_scheme.png}
\caption{Simple example of the correlation-inclusion coefficient $C^*$ between a modeled $D_S$ and observed distribution $D_O$. The indexes annotated $A$, $B$, $C$ and $D$, respectively correspond to the indexes where $D_O = 1$ while $D_S = 0$, $D_O = D_S = 0$, $D_O = D_S = 1$, and $D_O = 0$ while $D_S = 1$. We calculate the Pearson correlation coefficient between $D_O$ and $D_S$, but we anti-correlate the areas where the simulation predicts emissions that are not observed. To do so, we set at $-\sigma_O$ the null indexes of $D_O$ while $D_S$ occurrence is non-null. This correspond the index $D$. The other indexes $A$, $B$ and $C$ have theirs contributions unaltered. The will results here with a $C^* = 25\%$, while the Pearson's coefficient is $C = 42.9 \%$.}
\label{fig_si:grad}
\end{figure}

\begin{figure}
\centering
\noindent\includegraphics[width=\textwidth]{figures/annexe/all_gradient_epsilon_distribution.png}
\caption{Distributions of density gradient strength $||\nabla n_e||$ occurrences versus (a) the local plasma frequency $f_{pe}$, (b) the local upper-hybrid frequency $f_{uh}$ and (c) the second harmonic of the upper-hybrid frequency $2f_{uh}$ inside our simulation domain. 
The color scale indicates the number of grid cells for each value of $||\nabla n_e||$.
The dashed lines annotated $\epsilon_0$ and $\epsilon_9$ show respectively as examples, the $10^{th}$ and $90^{th}$ percentiles of the $||\nabla n_e||$ distribution at each frequency, annotated in the study $\epsilon = 0\% - 10\%$ and $\epsilon = 90\% - 100\%$.}
\label{fig_si:grad}
\end{figure}

\begin{figure}
\centering
\noindent\includegraphics[width=\textwidth]{figures/annexe/merged_fpe_fce_fX_map.png}
\caption{Meridian maps of (a) the ratio between the plasma frequency $f_{pe}$ and the cyclotron frequency $f_{ce}$, (b) the X-mode cutoff frequency $f_{X} = \sqrt{f_{pe}^2 + (f_{ce}/2)^2} + f_{ce}/2$. The meridian plane here is where the jovicentric, jovicentrifugal and jovimagnetic equator are aligned (defined by the sysIII longitude $\varphi = 200.769^{\circ}$). The vertical axis is the cartesian coordinate $z$. The horizontal axis is $\rho = (x^2 + y^2)^{1/2}$ the radial cylindrical coordinate in the meridian plane. The color maps have (0.1 $R_J$)$^2$ pixels. Jupiter's disk is displayed in black, the dotted circles around it mark radial distance by 1 $R_j$ steps, the dotted straight lines mark jovicentric latitudes by $10^\circ$ steps, and the bold dotted line is the jovicentric equator, aligned with the centrifugal and magnetic equator in this plane.}
\label{fig:plasma_map}
\end{figure}